\documentclass{IEEEoj}

\usepackage{cite}
\usepackage{amsmath,amssymb,amsfonts,array,multirow}
\usepackage{algorithmic}
\usepackage{graphicx,color}
\usepackage{eurosym}
\usepackage{textcomp}
\usepackage{tikz}
\usetikzlibrary{arrows.meta,positioning,calc}

\def\BibTeX{{\rm B\kern-.05em{\sc i\kern-.025em b}\kern-.08em
    T\kern-.1667em\lower.7ex\hbox{E}\kern-.125emX}}
\AtBeginDocument{\definecolor{ojcolor}{cmyk}{0.93,0.59,0.15,0.02}}
\def\OJlogo{\vspace{-4pt}\hskip-4pt\includegraphics[height=18pt]{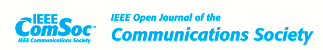}}
\begin{document}

\title{KNOWLEDGE GRAPH-BASED APPROACH FOR SUSTAINABLE 6G END-TO-END SYSTEM DESIGN}

\author{Akshay Jain\IEEEauthorrefmark{1} \IEEEmembership{(Senior Member, IEEE)}, Sylvaine Kerboeuf \IEEEauthorrefmark{2},  Sokratis Barmpounakis \IEEEauthorrefmark{3}, Crist\'{o}bal Vinagre Z.\IEEEauthorrefmark{4}, Stefan Wendt \IEEEauthorrefmark{5}, Dinh Thai Bui \IEEEauthorrefmark{2}, Pol Alemany\IEEEauthorrefmark{7}, Riccardo Nicolicchia\IEEEauthorrefmark{8}, Jos\'{e} Mar\'{i}a Jorquera Valero\IEEEauthorrefmark{9}, Dani Korpi\IEEEauthorrefmark{1}, Mohammad Hossein Moghaddam\IEEEauthorrefmark{10} \IEEEmembership{(Member, IEEE)}, Mikko A. Uusitalo\IEEEauthorrefmark{1} \IEEEmembership{(Senior Member, IEEE)}, Patrik Rugeland\IEEEauthorrefmark{6}, Abdelkader Outtagarts  \IEEEauthorrefmark{2} \IEEEmembership{(Senior Member, IEEE)}, Karthik Upadhya\IEEEauthorrefmark{1}, Panagiotis Demestichas\IEEEauthorrefmark{4}, Raul Muñoz\IEEEauthorrefmark{7}\IEEEmembership{(Senior Member, IEEE)}, Manuel Gil P\'{e}rez\IEEEauthorrefmark{9}, Daniel Adanza\IEEEauthorrefmark{7}, Ricard Vilalta\IEEEauthorrefmark{7}\IEEEmembership{(Senior Member, IEEE)}}
\affil{Nokia Bell Labs, Espoo, Finland}
\affil{Nokia France, Massy, France}
\affil{WINGS ICT Solutions SA, Greece}
\affil{TNO, The Hague, The Netherlands}
\affil{Orange Innovation, Ch\^{a}tillon, France}
\affil{Ericsson, Kista, Sweden}
\affil{Centre Tecnològic de Telecomunicacions de Catalunya (CTTC-CERCA), Castelldefels, Spain}
\affil{Telefónica Innovación Digital, Madrid, Spain}
\affil{Universidad de Murcia, Murcia, Spain}
\affil{Qamcom research and technology AB, Gothenburg, Sweden}
\corresp{CORRESPONDING AUTHOR: Akshay Jain (e-mail: akshay.2.jain@nokia-bell-labs.com).}
\authornote{This work was mainly supported by the Hexa-X-II Project, funded by the Smart Networks and Services Joint Undertaking (SNS JU) through the Horizon Europe Research and Innovation Programme under Grant 101095759, and was partially funded by the SHINE project (PID2024-159781OB-I00), funded by MICIU/AEI/10.13039/501100011033 and by ERDF/EU. }
\markboth{Knowledge graph-based approach for Sustainable 6G End-to-End System Design}{Jain \textit{et al.}}

\begin{abstract}
Previous generations of cellular communication, such as 5G, have been designed with the objective of improving key performance indicators (KPIs) such as throughput, latency, etc. However, to meet the evolving KPI demands and the ambitious sustainability targets for the Information and Communication Technology (ICT) industry, 6G will need to be designed differently. 6G will need to consider both the performance and sustainability targets for the various use cases it will serve. In addition, 6G will have various candidate technological enablers, making the design space of the system even more complex. Furthermore, due to the subjective nature of sustainability indicators, especially social sustainability, the literature still lacks clear methods to link them with technical enablers and 6G system design. Hence, in this article a novel method for 6G end-to-end (E2E) system design based on Knowledge graphs (KG) has been introduced. It considers as its input: the use case KPIs, use case sustainability requirements expressed as Key Values (KV) and KV Indicators (KVIs), the ability of the technological enablers to satisfy these KPIs and KVIs, the 6G system design principles defined in Hexa-X-II project, the maturity of a technological enabler and the dependencies between the various enablers. The KG method also introduces a novel approach for determining the key values addressed by a technological enabler. The effectiveness of the KG method was demonstrated by its application in designing the 6G E2E system for the cooperating mobile robot use case defined in the Hexa-X-II project, where 82 enablers were selected. Lastly, results from proof-of-concept demonstrations for a subset of the selected enablers have also been provided, which reinforce the efficacy of the KG method for designing a sustainable 6G system.          
\end{abstract}


\maketitle

\section{INTRODUCTION} \label{sec1}
\IEEEPARstart{P}{revious} generations of mobile systems were designed, optimized, and deployed to meet certain technical key performance indicators (KPI) such as latency and especially bitrate. However, early in the 6G research, it was identified that this technical focus needs to be broadened to be able to address the increasing challenges in our societies. The European flagship project Hexa-X introduced the concept of key value indicators (KVI) to 6G \cite{hexa_D11}.  These are intended to quantify the potential impact that the 6G system, use cases, and technologies have on sustainability. These KVIs would then be used to estimate and evaluate different technology and design choices based on their benefits and drawbacks in relation to their fulfillment of the traditional KPIs. \textcolor{black}{Where relevant, use case requirement ranges, enabler metadata, and PoC evaluation results are drawn from Hexa-X-II project deliverables and experiments. The proposed KG-based formulation and KPI/KVI-driven pruning methodology builds on these inputs to provide a generalized and reproducible 6G E2E system design procedure.
Additionally in this paper, a use case denotes a target 6G service scenario and its requirement profile, such as throughput, latency, trust, etc. An enabler denotes a technical capability, network function, or architectural component that contributes to fulfilling those requirements in an end-to-end system design. }\textcolor{black}{For example, in the cooperative mobile robots use case, stringent latency, reliability, and availability requirements translate into KPIs such as latency and service availability \cite{hexa2_D12,hexa2_D14}. Enablers such as AI-native air interface, intent-based management, and integration fabrics can be selected as technical means to satisfy these KPIs, while also contributing to sustainability-oriented KVIs such as reduced downtime, improved productivity, and improved trustworthiness within the proposed KG-based design workflow.}

However, the integration of sustainability values into the design of the 6G system presents significant challenges and requires fundamental changes in how we approach the 6G design. Notably, addressing the complexity of sustainability is a first challenge. Sustainability is multidimensional and encompasses a wide range of aspects, including social, economic, and environmental considerations. Integrating these diverse aspects into the design requires a holistic approach that considers the potential impact of each design decision across all dimensions of sustainability. Another challenge is to find a way to connect the desired outcomes in terms of sustainability to the design of the next generation of mobile technologies. The current and standard approach to translate requirements into indicators is using KPIs, which are measurable and objective. However, values are subjective, context-dependent, and difficult to measure. This creates a complex obstacle to translate them into actionable design requirements. Consequently, a novel framework is needed that effectively bridges the gap between the subjective values and concrete system requirements. 

Furthermore, deriving technical enablers, which are essential for the design of the entire 6G E2E system, is crucial. The vast amount of technologies underpinning the 6G system necessitates a systematic approach toward selecting the most impactful enablers. This involves identifying technologies that contribute significantly to the achievement of both the desired performance and sustainability goals, thus optimizing the efficiency and positive impact of the 6G E2E system.  In addition, it is necessary to assess the overall consistency of operations between the different components of the 6G system and to verify the ability to achieve the KPIs and KVIs targeted for a given use case. This requires a comprehensive approach to ensure that the system can meet the targeted KPIs and KVIs for a given use case, guaranteeing both performance and sustainability.

\textcolor{black}{We adopt a novel and comprehensive design framework to address the above challenges} to integrate an ever-growing set of technologies and services that the 6G system should offer to meet the needs of society in 2030 and beyond. This framework was first introduced in a previous work \cite{SK2024}. In this work, we delve deeper into the framework and also present an improvement over the previous approach \cite{hexa2_D23}. Specifically, a knowledge graph method and graph pruning approach is proposed towards the selection of enablers, while utilizing the defined objectives and the constraints. Based on this approach, distinct strategies or design choices based on the pursued values can be achieved. Concretely, the objective here is to optimize the 6G system considering all sustainability domains. 

Specifically, this work provides several key contributions:
\begin{itemize}
    \item It provides a concrete approach that connects the desired outcomes in terms of sustainability to 6G design. 
    \item It provides an analysis of the key values, which are then used to derive some KVIs. These are further exploited for deriving some tangible technical requirements for the 6G system design.
    \item A knowledge graph-based framework is proposed to select the technological enablers. This is done by taking into account the key requirements, design principles, and the sustainability ambitions for the 6G system, given a use case. It provides a method to assess the overall consistency of operations between the different components of the 6G system and to verify the ability to fulfill the KPIs and KVIs targeted for the use case under consideration. 
    \item To exemplify the KG based approach, the paper demonstrates the selection of technical enablers of the 6G system in a collaborative robot use case. 
    \item A performance and sustainability analysis of the selected enablers is provided based on results obtained from proof-of-concepts carried out within Hexa-X-II \cite{hexa_D26} project. 
\end{itemize}
The rest of this paper is organized as follows: Section \ref{sec2} builds a bridge from the sustainability values towards the 6G system design, wherein the state-of-the-art approaches to design 6G system are also briefly discussed. Section \ref{sec3} introduces the proposed knowledge graph-based framework for the KPI and KVI based 6G system design. Section \ref{sec4} presents the implementation of the Knowledge graph and graph pruning method in 6G system design for the cooperating mobile robots use case. Subsequently, in Section \ref{sec5} a performance and sustainability analysis of a subset of the enablers selected by the KG framework for the 6G E2E system has been presented. The paper is then concluded in Section \ref{sec6}. 

\section{FROM SUSTAINABILITY VALUES TO 6G DESIGN IMPACTS}\label{sec2}
In 2023, ITU released its framework for the IMT-2030 (i.e., 6G) which for the first time included sustainability as a target, thus acknowledging that future communication networks are supposed to deliver superior performance in a sustainable way \cite{ITUR2160}. In the context of mobile communications, sustainability can be divided into two main dimensions:
\begin{itemize}
    \item \textbf{Sustainable 6G:} It refers to ensuring that the sixth generation of mobile communications is sustainable \cite{hexa_D11, hexa2_D11}
    \item \textbf{6G for Sustainability:} It refers to how 6G can enable and enhance sustainability of the society at large \cite{hexa_D11, hexa2_D11}
\end{itemize}
For both of these dimensions, the Hexa-X-II project has considered the impact of sustainability in the form of three pillars, ie, environmental, social and economic sustainability \cite{hexa2_D11}. In particular, \cite{hexa2_D12,hexa2_D22,hexa2_D23} elaborate on the numerous ways in which 6G can not only comply with, but also promote these three fundamental pillars.

\textcolor{black}{Reference \cite{hexa2_D14} defines a set of values, named high-level human and planetary goals (HPGs), spanning the three sustainability pillars, and then evaluates how 6G use cases impact these HPGs.} \textcolor{black}{Because the HPGs span all three sustainability pillars, the evaluation also provides an impact analysis of use cases across these pillars.} The outcome of this evaluation process is defined as a ‘Key Value’ (KV). \textcolor{black}{A Key Value (KV) captures the impact of applying a use case on the HPGs}. To exemplify the latter, deploying connectivity in a previously unconnected rural area requires energy and materials to build and operate the infrastructure and devices deployed, impacting the environment. From a social perspective, it can help bridge the digital inclusion gap and bring better educational and healthcare possibilities. However, it can also present challenges in digital literacy. From an economic perspective, it can grant access to a wider range of products and services for people, but also empower local businesses. However, this also has to be achieved in a cost-efficient way for the network operators.

Next, to evaluate a positive or negative change in these values, the KVs need to be measured. Consequently, a set of indicators known as the KVIs are defined which can, directly or indirectly, measure a KV. It must be stated here that, as a use case is designed to interact within an ecosystem, many KVs (and consequently KVIs) fall beyond the reach of what can be influenced by 6G design. These are called Use Case KVIs, and encompass  first or higher order effects of the application of a use case \cite{hexa2_D12,hexa2_D14}. On the other hand, those KVIs which can be influenced by technical design are called Enabler KVIs \cite{hexa2_D12,hexa2_D14}. Note that, in this paper the focus is on the enabler KVIs and how they can be linked to the 6G system design itself, which can consequently help in achieving a net-positive impact on the overall use case level KVIs. Furthermore, it must be noted that enabler KVIs can be either mapped to an existing KPI, or to a new KPI, or stay as a subjectively determined enabler KVI \cite{hexa2_D14}. 

It must be stated that, there can be notable tradeoffs that need to be accounted for when doing the 6G E2E system design, while trying to adhere to the three sustainability pillars. For example, the environmental sustainability pillar will have energy reduction as one of its objectives. This will require considering the system-level integration of multiple technologies that provision energy reduction, such as energy neutral devices, AI-native air interface, etc. On the other hand, the societal sustainability pillar will require that the 6G system provisions sufficient level of trustworthiness. This will require the assessment of possible additional risks, for example in assuring that the AI-based components are explainable, interpretable, and fair to avoid any bias, as well as in evaluating how privacy and security of end-users are preserved. Additionally, the aforesaid assessment towards the impact on the environmental and social sustainability pillar requires considering the consistency in upholding the properties of energy efficiency and trustworthiness at an end-to-end level, i.e., at each sub-layer of the 6G system. 
\begin{table*}[htb]
    \centering
    \caption{\textcolor{black}{Differentiating aspects of current work as compared to state-of-the-art}}
    \begin{tabular}{|>{\centering\arraybackslash}m{1.5cm}|>{\centering\arraybackslash}m{1cm}|>{\centering\arraybackslash}m{5.5cm}|>{\centering\arraybackslash}m{1.5cm}|>{\centering\arraybackslash}m{1.5cm}|>{\centering\arraybackslash}m{2cm}|} \hline
         \textbf{\textcolor{black}{Paper}} & \textbf{\textcolor{black}{Uses KG}} & \textbf{\textcolor{black}{Focus and Methodology}} &  \textbf{\textcolor{black}{Use of KPIs}} & \textbf{\textcolor{black}{Use of KVIs}} & \textbf{\textcolor{black}{6G End-to-End system design}}\\ \hline
         
    \textcolor{black}{Current Work} & \textcolor{black}{Yes} & \textcolor{black}{Knowledge-graph-driven E2E system design methodology that maps use-case requirements, KPIs, values/KVIs, enablers, dependencies, and maturity constraints to systematically derive sustainable 6G design choices; validated on a cooperative mobile-robot use case} & \textcolor{black}{Explicit} & \textcolor{black}{Explicit} & \textcolor{black}{Yes (core contribution)} \\ \hline
    \textcolor{black}{\cite{hexa2_D23}} & \textcolor{black}{Yes} & \textcolor{black}{High-level E2E 6G system framework defining architectural layers, functional blocks, and design principles; provides a reference system design rather than a decision-making or selection methodology} & \textcolor{black}{No} & \textcolor{black}{No} & \textcolor{black}{High-level system design guidelines} \\ \hline
    \textcolor{black}{\cite{Yuan25}} & \textcolor{black}{Yes} & \textcolor{black}{KG-based multi-objective recommendation framework for air-interface configuration, empowered by a digital twin to adapt PHY/MAC parameters under service requirements}  & \textcolor{black}{Implicit (performance objectives)} & \textcolor{black}{No} & \textcolor{black}{No (component-level)} \\ \hline 
    \textcolor{black}{\cite{You2024WhenAM}} & \textcolor{black}{Yes (as an analysis step)} & \textcolor{black}{Uses knowledge graph analysis to identify critical datasets, followed by lightweight distributed AI and digital twins for real-time, sustainable 6G native intelligence} & \textcolor{black}{No (not central)} & \textcolor{black}{No} & \textcolor{black}{Partial (architectural AI framework, not E2E design method)} \\ \hline
    \textcolor{black}{\cite{merluzzi2026sustainable6gholisticview}} & \textcolor{black}{No} & \textcolor{black}{Conceptual, holistic analysis of sustainability trade-offs and enabling technologies for 6G; focuses on research questions rather than design workflows} & \textcolor{black}{No} & \textcolor{black}{No} & \textcolor{black}{No} \\ \hline
    \textcolor{black}{\cite{kamran2025energyaware6gnetworkdesign}} & \textcolor{black}{No} & \textcolor{black}{Survey and taxonomy of energy-aware and sustainable 6G network design approaches} & \textcolor{black}{Indirect (surveyed metrics)} & \textcolor{black}{No} & \textcolor{black}{No} \\ \hline
    \textcolor{black}{\cite{ahmadi2025sustainability6gbeyondchallenges}} & \textcolor{black}{No} & \textcolor{black}{Architectural and standards-oriented discussion of Open RAN as an enabler for sustainable 6G} & \textcolor{black}{No} & \textcolor{black}{No} & \textcolor{black}{Partial (architecture-level, not design methodology)} \\ \hline
        
    \end{tabular}
    \label{tab:reflist}
\end{table*}
Based on this example and the preceding background, without loss of generality it can be stated that the 6G system design space is multidimensional and complex. In \cite{hexa2_D23} an initial approach towards E2E 6G system design using a knowledge graph approach was proposed. However, the approach does not consider the Use case requirements, i.e., KPIs and KVIs, the enabler level KPIs and KVIs, in addition to the various other enabler features for the final enabler selection. Moreover, in \cite{hexa2_D23} a final alignment with the use case level KVs and KVIs has not been done. \textcolor{black}{Next, in \cite{Yuan25} knowledge graph (KG) is utilized to represent relationships between air interface configurations, channel behavior, and service requirements. Subsequently, digital twin and KG reasoning to recommend configurations that balance multiple performance metrics have been employed. On the other hand, \cite{Yuan25} does not provision not a full system design solution and it does not include any sustainability metrics. The authors in \cite{You2024WhenAM} utilize the KG based approach to look at the problem of energy efficient integration of AI into 6G networks. However, the paper does not delve into end-to-end system design and the associated trade-offs. In \cite{merluzzi2026sustainable6gholisticview}, a broad analysis of sustainability trade-offs and research questions for 6G has been presented. Specifically, general sustainability considerations have been mapped into technical themes that need innovation and standardization. Notably, this paper doesn’t provide a formal system design tool or specific methodology to optimize across KPIs and KVIs. The work in \cite{kamran2025energyaware6gnetworkdesign} focuses on energy efficiency and sustainability challenges and opportunities in network design at a high level. It does not propose a formal system design methodology nor uses a graph-based representation to jointly optimize KPIs and broader sustainability metrics. Lastly, in \cite{ahmadi2025sustainability6gbeyondchallenges} architectural sustainability challenges and opportunities with AI/ML and RAN softwarization have been discussed. It does not present a generalized system design framework nor includes a mechanism to tie performance targets with sustainability KPIs/KVIs formally.} 

Consequently, in this paper the authors build upon their initial work on Knowledge graph approach \cite{hexa2_D23} and evolve it to address the aforementioned gaps in the overall state-of-the-art. This provides a path forward in traversing such complex system design domain in an objective manner. \textcolor{black}{Specifically, as compared to the state-of-the-art, this paper:}

\begin{itemize}
    \item \textcolor{black}{Introduces the use case requirements, and utilizes them for 6G system design via Knowledge graphs. Specifically, concrete KPI and KVI requirements corresponding the use case have been utilized in this paper, while the state-of-the-art does not consider them.}
    \item \textcolor{black}{Performs a mapping from KVIs to 6G system enablers, thus allowing an improved evaluation of the enablers suitability towards satisfying sustainability goals set for the 6G system}
    \item \textcolor{black}{Provisions a proof-of-concept based analysis of the selected enablers, subsequently reinforcing the efficacy of the Knowledge graph based 6G system design process.}
\end{itemize}

\textcolor{black}{Table \ref{tab:reflist} presents a summary of the aspects that differentiate this paper as compared to the state-of-the-art.}
\begin{figure*}[!htb]
    \centering
    \includegraphics[width=1.7\columnwidth]{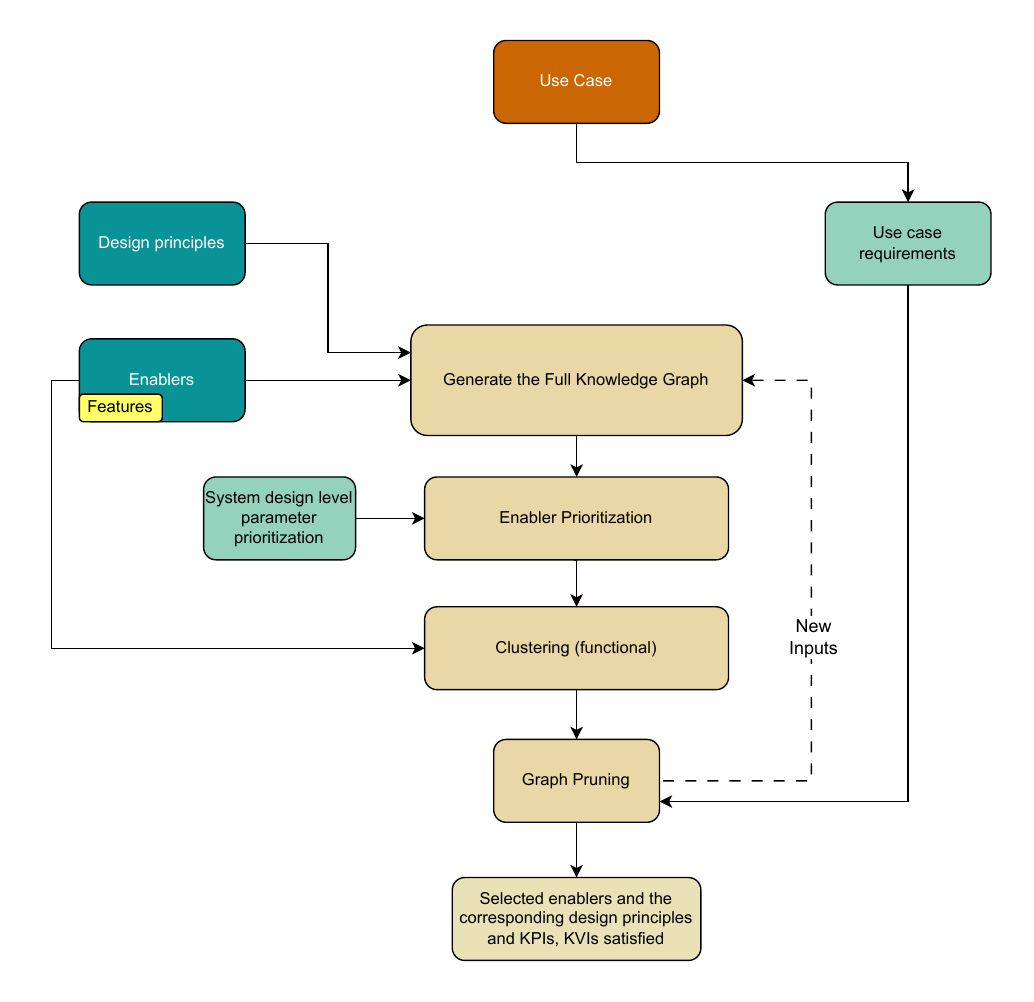}
    \caption{Knowledge Graph framework for 6G E2E system design}
    \label{fig:framework}
\end{figure*}
\section{KNOWLEDGE GRAPH FRAMEWORK FOR 6G END-TO-END SYSTEM DESIGN} \label{sec3}
Given the large number of enablers that will be explored during the 6G standardization and lifetime of the 6G system design, as well as their corresponding dependencies and features (KPIs, KVIs, etc.), a mechanism that can consider these complex relationships is required. Consequently, in this paper a framework is proposed that aims to provide a suitable set of enablers by considering a given set of requirements jointly. Such an approach allows one to visualize and evaluate various possible combinations of enablers, which can then be used toward designing and deploying the 6G system. \textcolor{black}{Furthermore, and as stated in Section \ref{sec2}, as compared to the state-of-the-art this work utilizes the use case requirements, i.e., KPIs and KVIs, performs a mapping from KVs and KVIs to the candidate enablers for the 6G system, and provisions a PoC based analysis of the selected enablers with respect to their capability to fulfill 6G sustainability requirements.}

The proposed KG framework, illustrated in Fig. \ref{fig:framework}, formulates the 6G E2E system design problem in the form of a graph construction and pruning problem. \textcolor{black}{Specifically, consider a graph $G = (\mathcal{V},\mathcal{E})$, where $\mathcal{V}$ represents the nodes and $\mathcal{E}$ the edges of the graph. For the 6G E2E system design, $G$ will represent a collection of nodes $\mathcal{V}$ representing enablers and design principles. These nodes $\mathcal{V}$ will have certain features such as maturity, KPIs, KVIs, etc., represented as a feature vector. Furthermore, they are connected to each other via edges $\mathcal{E}$ that capture the dependencies between enablers as well as how they fulfill design principles.} This KG-based approach allows capturing the dependencies between enablers, enabler features, design principles and use case requirements. Next, the edges of the full KG are of two types. These are elaborated as follows: 
\begin{itemize}
    \item \textbf{Enabler-Enabler:} Dependency of an enabler on another enabler mandates the presence of an edge between them in the KG
    \item \textbf{Enabler-Design principle:} If an enabler fulfills a design principle, then an edge is created between the enabler and the design principle 
\end{itemize}
\textcolor{black}{Note that, the full KG is an undirected graph wherein the edges do not have any direction.} \textcolor{black}{Given this full KG $G$, We seek a subset of enablers maximizing KPI/KVI alignment subject to the established constraints. Hence, the proposed method considers system-design parameter priorities, enabler functionalities, and use-case requirements}, and executes the following steps in order:
\begin{itemize}
    \item \textbf{Prioritization step:} This step prioritizes the enablers and their dependencies that are essential for the 6G E2E system design. This is based on the system design level parameters that are established as priorities. For example, if one of the criteria for the 6G system design is to ensure smooth migration from 5G to 6G, then the enablers that are critical for migration will be prioritized. \textcolor{black}{Concretely, The prioritization step does not down-select enablers. It assigns a categorical label, for example for migration $\mathrm{mig}_b\in\{-1,0,1\}$ (Eq.~(4)), that is used downstream to bias pruning against removing high-priority enablers.} 
    \item \textbf{Clustering:} Certain enablers might provide the same level of functionality, or they may achieve similar objectives. For example, one of the functionalities of the radio link control (RLC) is to perform retransmissions in the event of the loss of a service data unit (SDU). Similarly, if a code block group (CBG) or the complete transport block (TB) at the medium access control (MAC) layer is lost, then retransmissions are performed. Since, both the MAC and RLC enablers offer similar functionalities, they can be clustered together.  Hence, the proposed method offers the capability to cluster enablers based on functionality, which can then be used as a basis of selecting the best choice within the cluster of enablers
    \item \textbf{Graph Pruning:} This step considers the use case requirements and then it performs elimination of the unessential nodes, i.e., enablers and their corresponding dependencies, and edges from the KG so as to satisfy the constraints and objectives imposed by the use case requirements. 
\end{itemize}

In case new enablers or new dependencies of enablers, i.e., edges, are detected, the complete process is repeated after the graph pruning step. Lastly, the output of this framework is a pruned graph which is a representation of the set of selected enablers for the 6G E2E system given a use case and its corresponding requirements. Notably, the framework provisions the capability to execute the 6G system design for each use case independently, thus respecting the fact that each use case might have vastly different requirements which may demand a different set of enablers.

\section{6G END-TO-END SYSTEM DESIGN FOR THE COOPERATING MOBILE ROBOTS USE CASE} \label{sec4}
In this section, the KG method is employed for 6G E2E system design for the Cooperating Mobile Robots use case. Note that, Cooperating Mobile Robots is the representative use case of the Collaborative Robots use case family as identified in the Hexa-X-II project \cite{hexa2_D12,hexa2_D14}. \textcolor{black}{The use case is selected here as a representative industrial 6G scenario with stringent end-to-end requirements (e.g., latency, reliability, availability) and trust/efficiency considerations, making it well-suited to demonstrate KPI/KVI-driven enabler selection and pruning.}
\textcolor{black}{Consequently, it demonstrates how the proposed KG-based methodology extends beyond existing framework-level system designs (e.g., \cite{hexa2_D23}) by enabling requirement (KPI and KVI)-driven pruning and selection of design enablers, rather than prescribing a fixed architectural blueprint. This requirement-driven pruning is central to the paper’s novelty and is not addressed in our prior work \cite{hexa2_D23}.}

\subsection{Cooperating Mobile Robots Use case description} \label{sec4a}
Cooperating Mobile Robots (COBOTS) is the representative use case of the Collaborative Robots use case family \cite{hexa2_D12,hexa2_D14}. At the heart of this use case are autonomous robots that can move, sense their surroundings, and collaboratively perform productive tasks. These robots communicate with each other, other machines, and nearby humans to execute individual tasks that contribute to a shared cooperative objective. The purpose of this communication is to ensure safety and cooperation, enabling a group of robots to accomplish tasks beyond their individual capabilities and allowing each robot to perceive its environment beyond its immediate vicinity.

Effective cooperation hinges on reliable and fast exchange of relevant information between robots. Concretely, the deployment of cooperating mobile robots in industrial environments requires a network architecture that supports highly localized, task-specific interactions. \textcolor{black}{Manufacturing operations are high-stakes and even minor disruptions can result in significant financial and material losses.} To mitigate these risks, communication must meet strict reliability and low-latency requirements, particularly in safety-critical applications where human-robot interaction is involved. Hence, in Section \ref{sec4}-\ref{sec4b} the KPI and KVI requirements for the cooperating mobile robots use case have been discussed in detail. 


\subsection{KPI and KV/KVI requirements for COBOTS use case} \label{sec4b}
\begin{table*}[htb]
    \centering
    \caption{KPIs and corresponding target ranges for the Cooperating mobile robots use case}
    \begin{tabular}{|>{\centering\arraybackslash}p{4cm}|>{\centering\arraybackslash}m{2cm}|p{10cm}|} \hline
         \textbf{KPI} & \textbf{Target Range} & \hspace{3.8cm}\textbf{Comments/Justification}  \\ \hline
         Industrial Communication/ control traffic data rate [Mb/s] & $\leq 2.5$ DL and UL & Symmetric DL/UL data rates: UL=DL. The data rate might be rather low depending on the transfer interval and payload size \\ \hline
         Robot to campus data rate [Mb/s] & \hspace{0.25cm}$<10$ \newline $<250$ & Timely transfer of (raw) sensor data, video, immersive XR, AI/ML traffic, Digital Twin data, etc. The data rate might be significantly higher depending on the type of data exchanged with the campus network (e.g. video stream) \\ \hline
         Connection density [devices/$m^2$] & $\leq 0.5$ & Local concentration of mobile robots depends on the scenario, e.g., cooperative carrying assuming 8 robots in a space of 4 m x 4 m carrying an object \\ \hline
         Mobility [km/h] & $<20$ & Slow vehicular \\ \hline
         6G E2E latency (one way) D2D latency [ms] &\hspace{0.25cm} $\leq 5$ \newline $\geq 0.8$  & Exchange of coordination messages and control messages up to 600 times per second. Note, this results in a transfer interval of 1.66 ms. E2E latency limit is set to at most half that interval [22.104]. This provides enough margin for ARQ. \\ \hline
         6G E2E latency (one way) Robot to campus network [ms] & 1 - 10 & If timely response expected (e.g. digital twin data) Not critical, if no real-time response is expected (e.g. AI/ML model exchange) \\ \hline
         Max/min transfer interval [ms] & \hspace{0.25cm}$\leq 10$ \newline $\geq 1.66$ & Deterministic periodic communication with very low latencies in collaborative robotic use cases \\ \hline
         Message error rate & $< 10^{-5} - 10^{-7}$ & Message losses lead to interruptions in the collaboration process and may entail financial and material loss (damaged carried object). Application-side safety net mechanisms like “survival time” or “grace period” are employed to compensate for occasional packet losses and delays at the link level. Selected applications may have an even more strict packet reliability requirement of up to 99.999999\%, which is equivalent to a message error rate of 10-9.  \\ \hline
         Message size [byte] & 250-500 & Applications are putting more and more information into the control data, 500 bytes with localization information \\ \hline
         Coverage [\%] & 100 & Localized nature of a joint task makes local ad hoc connectivity favourable. \\ \hline
         Positioning accuracy [m] & \hspace{0.15cm}$< 0.1$ (fine) \newline $<1$ (coarse) & Tasks such as environment mapping, robot navigation, and inventory management require fine positioning. Tasks like robot localization need coarse positioning. \\ \hline
         Human presence detection [m] & 0.4 & Identifying human presence is very important: 40 cm resolution within 1 m of the machine, velocity. (Scenarios autonomous construction, smart workshop) \\ \hline
         Sensing-related capabilities [Y/N] & Yes & Robots and cobots depend on capturing the environmental context. Network-integrated sensing may complement or replace dedicated onboard sensors. Efficient transport of data/information from connected external sensors is likely needed. \\ \hline
         AI/ML-related capabilities [Y/N] & Yes & Robots and cobots depend on advanced machine learning. Execution may be embedded in the device and/or offloaded at a local edge and/or provided by the network as an over-the-top service. \\ \hline
    \end{tabular}
    \label{tab:Tab1}
\end{table*}

The KPI and KVI requirements for the cooperating mobile robots use case have been detailed in Tables \ref{tab:Tab1} and \ref{tab:tab2}, respectively. \textcolor{black}{Note that.  these KPI and KVI requirements are taken from Hexa-X-II use case requirement definitions \cite{hexa2_D12,hexa2_D14} and used as input to the proposed KG-based method.} Specifically, in scenarios where ensuring seamless coordination among robots is paramount, delays or failures in message delivery can compromise operational safety and efficiency. On the other hand, as automation advances and machine autonomy increases, some applications may tolerate slightly relaxed reliability and latency constraints, particularly in non-critical tasks. These requirements have been clearly elaborated in Table \ref{tab:Tab1}, wherein a target range of 1-10 ms for the robot to campus 6G network latency has been determined.

A crucial aspect of this robotic ecosystem is mobility. Given that cooperating robots operate within dynamic environments, their sub-networks are often temporary and adaptive, allowing them to seamlessly form, dissolve, and migrate within a broader campus network. This nomadic behavior introduces challenges in maintaining seamless handovers, both within single industrial sites and across interconnected facilities. The ability to sustain low-latency, high-reliability communication during movement is key to ensuring uninterrupted robotic coordination and task execution. These have been captured as part of the 6G E2E latency, coverage, mobility support as well as message error rate KPIs listed in Table \ref{tab:Tab1}.

\begin{table*}[!htb]
    \centering
    \caption{KV and enabler KVI for Cooperating Robots}
    \renewcommand{\arraystretch}{1.1}
    \begin{tabular}{|>{\centering\arraybackslash}p{2cm}|p{6cm}|p{6cm}|>{\centering\arraybackslash}p{2cm}|} \hline
  
         \textbf{Sustainability Pillar} & \hspace{2.9cm}\textbf{KV} & \hspace{2.2cm}\textbf{Enabler KVI(s)} & \textbf{KVI category} \\ \hline
         \multirow{3}{*}{\vspace{-2.3cm} \textbf{Environmental}} & Resource efficiency: Functionalities may be provided by machines with less materials, energy, and waste generated & 
             Energy consumption per process \slash overall [MWh]
        & \multirow{2}{*}{\vspace{-1.8cm}\textbf{Energy}}\\ \cline{2-3}
         & Energy is consumed and materials are used to manufacture, deploy, and operate robots and associated services & Energy used in operations [MWh], Energy used in data transfer [MWh], Storage data efficiency [TB] & \\ \cline{2-4}
         & The disposal of machines and devices results in increased electronic waste & Life expectancy of robots [unit of time], number of virtualized functionalities & \textbf{Materials\slash Waste} \\ \hline
         \multirow{2}{*}{\vspace{-0.5cm}\textbf{Social}} &  Safer work environment leading to fewer injuries &  Number of injuries at work
         & \textbf{Safety} \\ \cline{2-4}
         &  Number of data leaks\slash breaches\slash cyber attacks with personal information compromised & Reliability, Trust and Security, Explainability
         & \textbf{Trustworthiness\slash Privacy\slash Security} \\ \hline
         \multirow{3}{*}{\vspace{-3.5cm}\textbf{Economic}} & Increased local and global productivity, cost efficiency and enhanced competitiveness from the use of the collaborative robots & Downtimes [hours], Non-conformance cost [\euro] & \textbf{Productivity\slash Efficiency}\\ \cline{2-4}
         & New business opportunities for old and new businesses from the use of collaborative robots (new markets around robots beyond the factory) & Average price of cobots [\euro] 
        & \multirow{3}{*}{\vspace{-2.2cm}\textbf{Costs}} \\ \cline{2-3}
         & High initial CAPEX\slash investment as a hurdle for starting a business with use of cobots, including building skill sets, buying equipment, etc. &  Initial investment required for setting up Collaborative Robots [\euro]
         & \\ \cline{2-3}
         & Higher maintenance costs compared to traditional robots & OPEX from equipment, maintenance, operation, service, training, etc. of CMR [\euro], Running CAPEX [\euro]
      &  \\ \hline
      \multicolumn{4}{c}{\textcolor{black}{Note: Use-case KVIs are not listed here because they represent higher-order impacts that depend on}} \\
      \multicolumn{4}{c}{\textcolor{black}{deployment, regulations, user behaviour, etc., which are beyond technical design; see \cite{hexa2_D12}, \cite{hexa2_D14}.}}
               
    \end{tabular}
   
    \label{tab:tab2}
\end{table*}
Beyond connectivity, precise sensing and positioning capabilities are essential for navigation, task execution, and safety. Tasks such as environment mapping, robot localization, and inventory management require different levels of positioning accuracy, with fine precision being necessary for highly coordinated movements, while coarser localization may suffice for broader tracking applications. The integration of network-assisted sensing enhances perception, complementing onboard sensors with additional data from the 6G infrastructure. By fusing device-based sensor data with network-integrated sensing technologies, robots can achieve greater situational awareness, improving navigation and decision-making in dynamic environments. Specifically, in Table \ref{tab:Tab1} fine and coarse positioning accuracy requirements have been specified. Furthermore, the resolution accuracy for detecting humans has also been specified in Table \ref{tab:Tab1}.

\begin{figure*}[!htb]
    \centering
    \includegraphics[width=1.7\columnwidth]{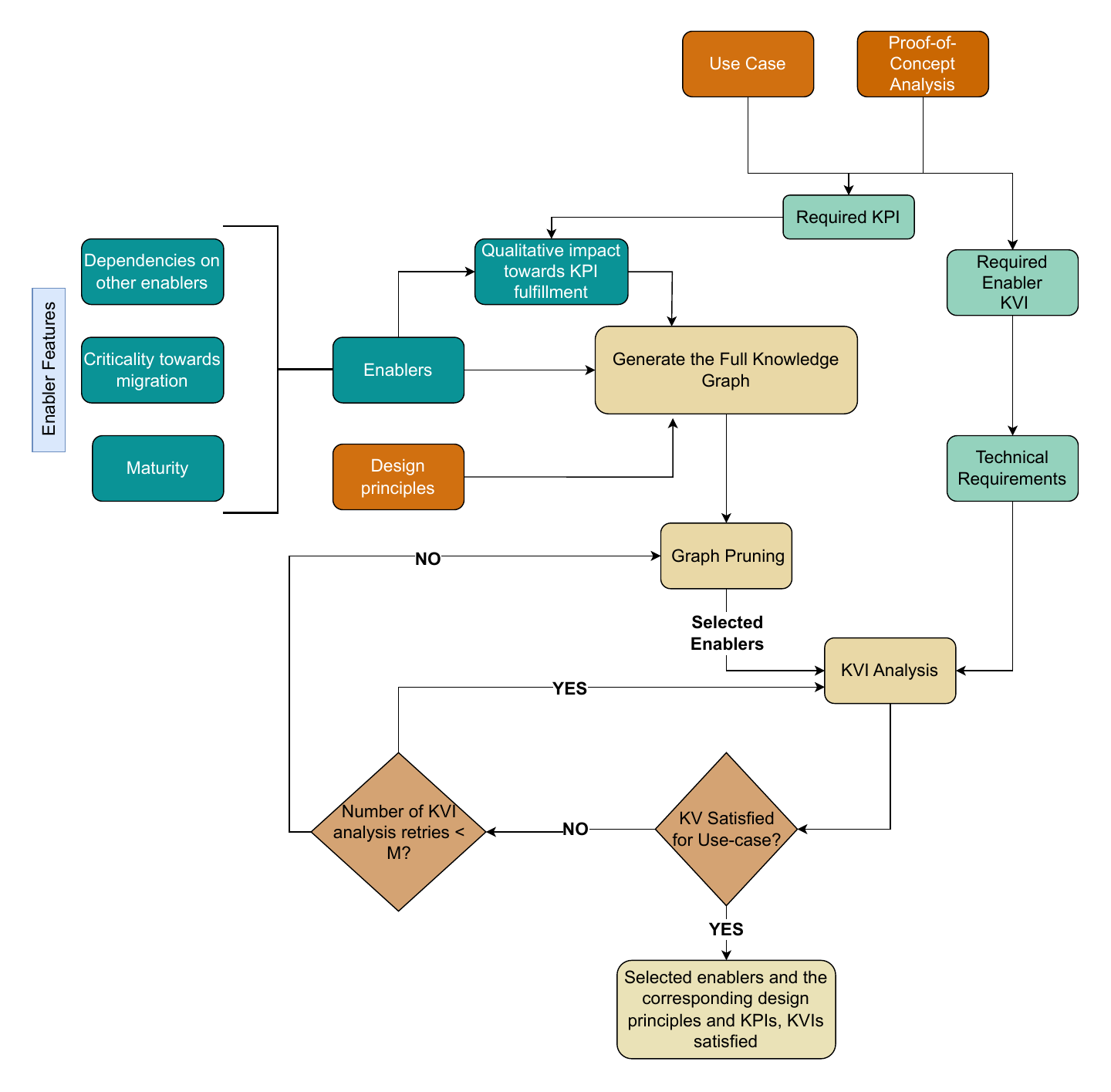}
    \caption{Knowledge Graph based enabler selection method implementation for Cooperating Mobile Robots Use case}
    \label{fig:process}
\end{figure*}

Next, AI/ML-driven capabilities play a fundamental role in enhancing robotic intelligence and coordination. Machine learning algorithms enable real-time data processing, allowing robots to identify patterns, detect anomalies, and optimize task execution autonomously. These AI-driven processes can be executed locally on the robot or offloaded to edge computing nodes, depending on the computational complexity and latency constraints. Efficient transport and management of AI/ML traffic ensure that multiple robots can learn and adapt collaboratively, further optimizing performance in highly automated industrial settings. Hence, the convergence of localized network structures, ultra-reliable low-latency communications, precise positioning, and AI-enhanced sensing forms the backbone of next-generation industrial robotics, fostering a highly efficient, adaptable, and intelligent automation ecosystem.

It is important to note that most KVs and KVIs fall outside the scope of what can be influenced by 6G. This is because they are directly related to the implementation of the use case, i.e. Use Case KVIs, which concerns the social sustainability domain especially. \textcolor{black}{In particular, these use case KVIs are typically second- or third-order outcomes (e.g., workplace safety practices or societal impact) that depend strongly on deployment context, operational processes, and human factors, and thus cannot be directly controlled through 6G technical design alone.} However, other KVIs can be influenced by the design and implementation of 6G, i.e. the enabler KVIs. We provide in Table \ref{tab:tab2} exclusively the KVs and enabler KVIs for the cooperating mobile robots use case \cite{hexa2_D12,hexa2_D14}. Specifically, the KVs and enabler KVIs for the three sustainability pillars given the cooperating robots use case have been listed in Table \ref{tab:tab2}. Furthermore, broad categories for the enabler KVIs, i.e., \textit{Energy}, \textit{Materials\slash Waste}, \textit{Safety}, \textit{Trustworthiness\slash Privacy\slash Security}, \textit{Productivity\slash Efficiency} and \textit{Costs}, have also been provided. 

The following sections utilize these KPIs and enabler KVIs to perform the 6G system design. 


\subsection{Enabler selection methodology} \label{sec4c}
Based on the KG method framework defined in Section \ref{sec3}, \textcolor{black}{we next provide a formal representation of the KG construction and the associated enabler selection problem. The formalization defines the node/edge sets, feature spaces, and the constrained selection objective that guides graph pruning. While the selection structure is formalized mathematically, the KVI contribution scores are instantiated using an explicit heuristic rubric, since standardized models for quantifying sustainability impact, particularly for social sustainability, remain limited and context-dependent.}

\textcolor{black}{Let the initial full knowledge graph be defined as}
\begin{equation}
\textcolor{black}{G = (\mathcal{V},\mathcal{E})}
\end{equation}
\textcolor{black}{where the node set is}
\begin{equation}
\textcolor{black}{\mathcal{V} = \mathcal{B} \cup \mathcal{D}}
\end{equation}
\textcolor{black}{Here, $\mathcal{B}$ denotes enabler nodes and $\mathcal{D}$ denotes design-principle nodes. The edge set is}
\begin{equation}
\textcolor{black}{\mathcal{E} = \mathcal{E}_{BB} \cup \mathcal{E}_{BD}}
\end{equation}
\textcolor{black}{where $\mathcal{E}_{BB}$ represents dependencies between enablers and $\mathcal{E}_{BD}$ represents the relation that an enabler fulfills a design principle.}

\textcolor{black}{Each enabler node $b\in \mathcal{B}$ is associated with a feature vector}
\begin{equation}
\textcolor{black}{\mathbf{h}_b = [\mathrm{p_b},\, \mathrm{hw_b},\, \mathrm{trl_b},\, \mathrm{std_b},\, \mathrm{mig_b},\, \mathbf{f}_b]}
\end{equation}
\textcolor{black}{where $\mathrm{p}_b\in\{-1,0,1\}$ denotes low/medium/high priority, $\mathrm{hw}_b\in\{0,1\}$ indicates whether new hardware entities are required, $\mathrm{trl}_b\in\{1,\dots,9\}$ is the technology readiness level (TRL), $\mathrm{std}_b\in\{0,1\}$ indicates whether a standards update is required, $\mathrm{mig}_b\in\{0,1\}$ indicates whether the enabler is essential for migration, and $\mathbf{f}_b$ is a multi-hot vector encoding functionality tags (e.g., localization, edge computing).}

\textcolor{black}{Edge features are represented as binary indicators:}
\begin{equation}
\textcolor{black}{e_{b,d} = \mathbb{I}\{b \rightarrow d\}, \quad \forall (b,d)\in\mathcal{E}_{BD}}
\end{equation}
\begin{equation}
\textcolor{black}{e_{a,b} = \mathbb{I}\{a \rightarrow b\}, \quad \forall (a,b)\in\mathcal{E}_{BB}}
\end{equation}

\textcolor{black}{where $\mathbb{I}\{\cdot\}\in\{0,1\}$ denotes the indicator function. To capture dependency integration difficulty, dependency edges may be weighted as}
\begin{equation}
\textcolor{black}{w^{dep}_{a,b}=e_{a,b}\cdot\omega_{a,b}, \quad 0\le \omega_{u,v} < 1, \forall (a,b) \in\mathcal{E}_{BB}}
\end{equation}
\textcolor{black}{where $\omega_{a,b}$ reflects the relative maturity of the interface between enablers $a$ and $b$.}

\textcolor{black}{Let $x_b\in\{0,1\}$ denote whether enabler $b\in\mathcal{B}$ is selected. The selection problem can be written as:}
\begin{equation}
\textcolor{black}{\max_{\mathbf{x}} \sum_{b\in\mathcal{B}} x_b \cdot \Big(\alpha S^{KPI}_b + \beta S^{KVI}_b + \gamma S^{DP}_b + \delta \mathrm{mig}_b\Big),}
\end{equation}
\textcolor{black}{subject to dependency and feasibility constraints such as:}
\begin{equation}
\textcolor{black}{x_a \le x_b \quad \forall (a,b)\in\mathcal{E}_{BB} \text{ where } e_{a,b}=1,}
\end{equation}
\begin{equation}
\textcolor{black}{\mathrm{trl}_b \ge \tau \quad \forall b\in\mathcal{B}\text{ with }x_b=1.}
\end{equation}

\textcolor{black}{Here, $S^{KPI}_b$ represents the KPI-alignment score derived from the qualitative impact assessment, $S^{DP}_b$ represents the design-principle alignment, and $S^{KVI}_b$ represents the sustainability/value alignment score derived from the KVI mapping. In this work, the KVI contribution score is instantiated via an explicit deterministic rubric:}
\begin{equation}
\textcolor{black}{S^{KVI}_b=\phi\!\left(\mathbf{h}_b,\, \{e_{b,d}\}_{d\in\mathcal{D}},\, \{e_{b,a}\}_{a\in\mathcal{B}}\right),}
\end{equation}
\textcolor{black}{which enables repeatable scoring and controlled pruning under the same set of assumptions and thresholds.Fig.~\ref{fig:toyKG_weights_vectors} shows a toy KG example illustrating node types, edge types, dependency weights, and the enabler feature vectors in Eq.~(4).
}

\begin{figure*}[t]
\centering
\begin{tikzpicture}[
  >=stealth,
  thick,
  font=\footnotesize,
  enabler/.style={draw, rounded corners, fill=blue!10, minimum width=3.0cm, minimum height=1.0cm, align=center},
  dp/.style={draw, rounded corners, fill=green!12, minimum width=3.2cm, minimum height=0.9cm, align=center},
  vecbox/.style={draw, rounded corners, fill=gray!6, inner sep=3pt, align=left},
  edgeDepa/.style={->, red},
  edgeDepb/.style={->, green!60!black},
  edgeDP/.style={->, green!60!black}
]

\node[enabler] (a) at (0,0) {Enabler $a$\\(e.g., AI-native air interface)};
\node[enabler] (b) at (7.2,0) {Enabler $b$\\(e.g., Integration fabric)};
\node[enabler] (c) at (7.2,-2.9) {Enabler $c$\\(e.g., LoTAF)};
\node[dp] (d) at (5.0,2.2) {Design principle $d$\\(e.g., modularization)};

\node[vecbox, anchor=east] (va) at ($(a.west)+(-0.9,1)$) {%
$\mathbf{h}_a=\left[\begin{array}{c}
\mathrm{p}_a \\[1pt]
\mathrm{hw}_a \\[1pt]
\mathrm{trl}_a \\[1pt]
\mathrm{std}_a \\[1pt]
\mathrm{mig}_a \\[1pt]
\mathbf{f}_a
\end{array}\right]$
};

\node[vecbox, anchor=west] (vb) at ($(b.east)+(0.9,1)$) {%
$\mathbf{h}_b=\left[\begin{array}{c}
\mathrm{p}_b \\[1pt]
\mathrm{hw}_b \\[1pt]
\mathrm{trl}_b \\[1pt]
\mathrm{std}_b \\[1pt]
\mathrm{mig}_b \\[1pt]
\mathbf{f}_b
\end{array}\right]$
};

\node[vecbox, anchor=west] (vc) at ($(c.east)+(0.9,0)$) {%
$\mathbf{h}_c=\left[\begin{array}{c}
\mathrm{p}_c \\[1pt]
\mathrm{hw}_c \\[1pt]
\mathrm{trl}_c \\[1pt]
\mathrm{std}_c \\[1pt]
\mathrm{mig}_c \\[1pt]
\mathbf{f}_c
\end{array}\right]$
};

\draw[edgeDepa] (a) -- node[midway, below=2pt] {$w^{dep}_{a,b}=0$} (b);

\draw[edgeDepb] (b) -- node[midway, left=2pt] {$w^{dep}_{b,c}=1$} (c);

\draw[edgeDP] (b) -- node[midway, right=2pt] {$w_{b,d}=1$} (d);

\node[vecbox, anchor=north west] (legend) at (-5.2,-1.8) {%
\textbf{Toy KG semantics:}\\
$\mathcal{E}_{BB}$: enabler dependency edge (weighted)\\
$\mathcal{E}_{BD}$: enabler supports design principle ($e_{b,d}\in\{0,1\}$)\\[2pt]
\textbf{Node features (Eq.~(4)):}\\
$\mathrm{p}$: prioritization,\ $\mathrm{hw}$: hardware need,\ $\mathrm{trl}$: maturity,\ $\mathrm{std}$: standards impact,\\
$\mathrm{mig}$: migration essentiality,\ $\mathbf{f}$: multi-hot functionality tags
};

\end{tikzpicture}
\caption{Toy KG example illustrating (i) weighted dependency edges between enablers and (ii) an enabler-to-design-principle relation, together with enabler feature vectors (Eq.~(4)) capturing prioritization, hardware impact, TRL, standards impact, migration essentiality, and functionality tags.}
\label{fig:toyKG_weights_vectors}
\end{figure*}

\textcolor{black}{While (1)–(11) provide a formal representation of the KG-based enabler selection problem, directly solving this formulation is challenging in practice due to the combinatorial nature of the search space and the context-dependent (partly subjective) instantiation of KVI contribution scores. Therefore, we operationalize the selection using the methodology shown in Fig.~\ref{fig:process}, which performs (i) qualitative KPI impact assessment, (ii) full KG generation, (iii) KPI-threshold-based graph pruning, (iv) mapping of the retained enablers to enabler KVIs followed by KVI analysis, and (v) iterative pragmatic considerations with updated pruning thresholds until the use case KVs are satisfied. In Fig. \ref{fig:process} firstly,} the use case requirements \cite{hexa2_D12,hexa2_D14} and proof-of-concept analysis \cite{hexa_D24} are utilized to derive the KPI and enabler KVI requirements, which have been presented in Tables \ref{tab:Tab1} and \ref{tab:tab2} of Section \ref{sec4}-\ref{sec4b}. Next, the required enablers along with their features are utilized to derive the qualitative impact of each enabler for every KPI requirement. Such a qualitative analysis is carried out because for a given service level agreement (SLA), an enabler in the radio might have different requirements as compared to another enabler in the Management and Orchestration domain, while both are necessary for ensuring the SLA. For example, to achieve resiliency in network operations, the radio domain enablers will need to provision low block error rates/message error rates, while the Management and Orchestration domain enablers will need to provision reduced reconfiguration time/service migration time. Hence, for a unified analysis and selection process, the impact that an enabler has towards satisfying a given KPI is characterized as being positive, negative or neutral.

\begin{figure*}[!htb]
    \centering
    \includegraphics[width=2\columnwidth]{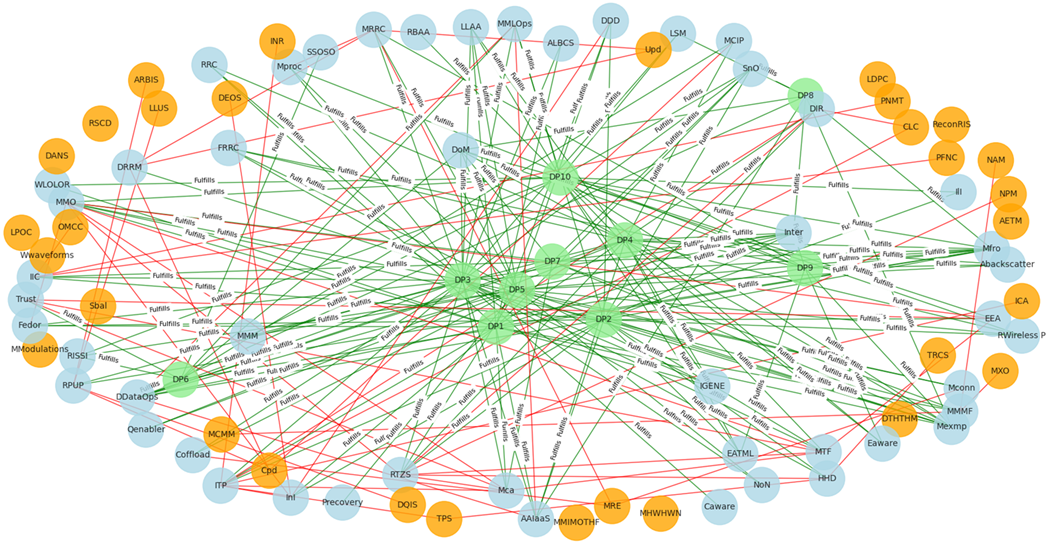}
    \caption{Full Knowledge Graph}
    \label{fig:fullKG}
\end{figure*}

Subsequently, as shown in Fig. \ref{fig:fullKG}, utilizing the enabler and its corresponding features, the design principles, and the qualitative impact of the enablers on the KPI requirements, the full KG is generated. In the full KG, the enablers are represented via blue and orange bubbles, and the design principles are represented via green bubbles. The enablers represented via blue bubbles are those that satisfy at least one design principle and hence have a green edge going towards at least one green bubble. \textcolor{black}{The enabler to design principle dependency is established through the meta-data collected within the Hexa-X-II project \cite{Jain2025}, which is defined and discussed in more detail in Section \ref{sec4}-\ref{sec4d}.} However, enablers represented via orange bubbles do not satisfy any design principles. Additionally, the dependency between enablers is captured via the red edges.

The full KG is then pruned, as presented in the methodology shown in Fig. \ref{fig:process}. The pruning process is done by selecting a threshold on the overall enabler KPI impact. This is elaborated further in Section \ref{sec4}-\ref{sec4e}. Next, the selected enablers from the graph pruning step are then mapped to the enabler KVIs. To perform this, the enabler KVIs obtained from the use case requirements, detailed in Table \ref{tab:tab2} Section \ref{sec4}-\ref{sec4b}, are first mapped to the technical requirements, which are presented in Table \ref{tab:tab3} Section \ref{sec4}-\ref{sec4d}. The enablers are then mapped to these technical requirements based on their capabilities to satisfy the said technical requirements. Consequently, the enablers are mapped to the enabler KVIs. 

Given the mapping between enablers and enabler KVIs, a KVI analysis is performed, the details of which are deferred to Section \ref{sec4}-\ref{sec4e}. At this stage, it is determined whether the set of selected enablers can satisfy the KVs determined for the use case under consideration. If so, then the set of selected enablers are finalized for the 6G E2E system. On the other hand, if the KVs are not satisfied then a round of pragmatic considerations is done. The pragmatic considerations step determines which of the enablers that have been removed at the graph pruning step need to be re-considered. Subsequently, the KVI analysis is repeated to verify the improvements and the fulfillment of the use case KVs. In case, even after $M$ repetitions of the pragmatic considerations and KVI analysis the KVs are not satisfied, the enabler selection process is repeated from the graph pruning step, as shown in Fig. \ref{fig:process}. It must be stated that, the repetition of the graph pruning step is done on the full KG with a new set of thresholds. \textcolor{black}{To facilitate interpretation of Fig.~\ref{fig:process}, Table~\ref{tab:kg_flow_steps} summarizes the main steps of the KG-based methodology and their underlying rationale.}

\begin{table}[t]
\caption{\textcolor{black}{Simple description of the KG-based enabler selection method}}
\label{tab:kg_flow_steps}
\centering
\renewcommand{\arraystretch}{1.1}
\begin{tabular}{|p{0.20\linewidth}|p{0.7\linewidth}|}
\hline
\textcolor{black}{\textbf{Step}} & \textcolor{black}{\textbf{Purpose and rationale (simple description)}}\\
\hline
\textcolor{black}{Use case requirement ingestion} & \textcolor{black}{Collect the use case KPI and KVI targets (inputs) to define what the system must achieve.}\\
\hline
\textcolor{black}{Qualitative KPI impact scoring} & \textcolor{black}{Normalize heterogeneous KPI interpretations across domains by assigning each enabler a positive/neutral/negative impact per KPI requirement.}\\
\hline
\textcolor{black}{Full KG generation} & \textcolor{black}{Build the initial graph capturing enablers, dependencies, and design-principle relations to enable systematic reasoning at E2E level.}\\
\hline
\textcolor{black}{KPI-threshold pruning} & \textcolor{black}{Remove low-impact enablers using an overall KPI impact threshold to reduce the search space while preserving KPI feasibility.}\\
\hline
\textcolor{black}{Enabler-to-KVI mapping} & \textcolor{black}{Translate KVs/KVIs into technical requirements and map enablers to the KVIs they support based on capability alignment.}\\
\hline
\textcolor{black}{KVI analysis} & \textcolor{black}{Verify whether the selected enabler set provides sufficient KVI coverage (i.e., sustainability/value alignment) for the use case.}\\
\hline
\textcolor{black}{Pragmatic considerations loop} & \textcolor{black}{If KVs are not satisfied, reintroduce specific removed enablers and/or adjust pruning thresholds, then repeat KVI analysis until convergence.}\\
\hline
\textcolor{black}{Final enabler set} & \textcolor{black}{Output a KPI-feasible and KVI-aligned enabler subset for 6G E2E system design.}\\
\hline
\end{tabular}
\end{table}


\subsection{Meta-data collection and transformation} \label{sec4d}
The enabler selection process, as presented in Section \ref{sec4}-\ref{sec4c}, firstly requires the collection of enabler meta-data. This meta-data is composed of the enabler feature information, as illustrated in Fig. \ref{fig:process}, which also includes the impact of the enabler on the KPIs. Specifically, the positive, negative or neutral impact of the enablers for achieving the KPIs provisioned in Table \ref{tab:Tab1} is also captured as part of the enabler features.

Subsequently, to construct the full KG, as discussed in Section \ref{sec4}-\ref{sec4c}, the collected meta-data needs to be transformed. It must be stated that in this article the enabler meta-data considered is taken from the open source repository provided by the Hexa-X-II project \cite{Jain2025}.  From the provisioned meta-data, in this article the inputs that have been used to construct the knowledge graph are maturity, criticality towards migration, KPIs (Table \ref{tab:Tab1}), design principles fulfilled, enabler dependencies and enabler KVIs (Table \ref{tab:tab2}). 

Each of the features considered is, if required, transformed to an objective scale. The maturity of an enabler, which is expressed as the technological readiness level (TRL) on a scale of 1–9 \cite{APRE_CDTI}, readily translates to an objective quantity that can be used for graph development, comparison and graph pruning processes. Next, the importance of an enabler towards migration is objectively expressed as 'yes' or 'no'. \textcolor{black}{This is transformed into an objective value by assigning a fixed weight greater than 1 for those enablers with a value of ``yes'', which in this article is set to 3 as a pragmatic design bias to prevent migration-essential enablers from being down-ranked during pruning. This value is not optimized and could equivalently be represented as a binary flag.} On the other hand, for those enablers with a value ``no'', a weight of 1 is used. This weighting scheme ensures that the enablers that are essential for migration are prioritized, in line with the evolutionary nature of 6G vis-\`a-vis 5G.

The meta-data for the impact of the enabler on the KPIs \cite{Jain2025} are encoded by the triple [$-1, 0, +1$], wherein $-1$ represents a negative impact, $0$ represents a neutral impact and $+1$ represents a positive impact. The overall KPI score for an enabler is computed as the sum of the impacts on all the KPIs. This is subsequently used in the graph pruning process. The fulfillment of design principles is captured through the edge weights, wherein the existence of an edge between an enabler and a design principle is given a weight $1$. On the other hand, there is no edge if an enabler does not fulfill a certain design principle. In addition, the dependencies between enablers are also captured via edges. However, we specify an edge weight of $0$ to distinguish these edges from those that are used to establish the relationship between enabler and design principles. Table \ref{tab:Tabfeat} summarizes the weights and meta-data transformations introduced above.  

\begin{table}[htb]
    \centering
    \renewcommand{\arraystretch}{1.1}
    \caption{Node and Edge feature values for Knowledge Graph pruning}
    \begin{tabular}{|p{6cm}|>{\centering\arraybackslash}p{1cm}|} \hline
         \hspace{1.5cm}\textbf{Node and Edge Features}& \textbf{Feature values}  \\ \hline
         Enabler - Design principle edge weight & 1 \\ \hline
         Enabler - Enabler edge weight & 0 \\ \hline
         Positive KPI impact & 1 \\ \hline
         Neutral KPI impact & 0 \\ \hline
         Negative KPI impact & -1 \\ \hline
         Node weight for enablers critical for migration & 3 \\ \hline
         Node weight for enablers not critical for migration & 1 \\ \hline
    \end{tabular}
    
    \label{tab:Tabfeat}
\end{table}

Next, for the enabler KVIs based assessment, the stated enabler KVI requirements are first distilled into technical requirements, after which it is determined if a given enabler satisfies those requirements. Based on the enabler to the technical requirement mapping, they are then mapped as being able to satisfy the corresponding enabler KVI requirement. A detailed description of the enabler KVI to technical requirement mapping is presented in Table \ref{tab:tab3}. These technical requirements, and hence, the enabler KVI requirements are then used to align the selected enablers, such that the final list of enablers satisfies both the KPI and KVI requirements for the use case under consideration. Note that the term "align" means that either more enablers may be added, or some enablers may be removed. The main reason for the aforementioned subjective analysis is because:
\begin{itemize}
    \item The entire 6G system will consist of a large number of enablers, located in the different stacks of the E2E system \cite{hexa2_D23}. Each of these enabler provisions a certain KPI and KVI value. Notably, these will be different at different stacks of the E2E system, given that the requirements at those stacks will be different to satisfy the use case KPI and enabler KVI requirements.
    \item Certain enablers are at a significantly low TRL level, for example TRL 1. For such enablers, the KPI and KVI assessments do not exist as they may still be under development.  
    \item Certain KVIs do not have concrete metrics yet. For instance, social sustainability metrics are hard to define and are usually subjective in nature. 
\end{itemize}

\begin{table}[htb]
    \centering
     
    \caption{KVIs to technical requirements mapping}
    \begin{tabular}{|p{4cm}|p{4cm}|} \hline 
 
        \hspace{1.2cm} \textbf{Enabler KVI(s)} & \hspace{1.2cm}\textbf{Requirements}  \\ \hline
            Energy consumption per process/overall [MWh], Energy used in operation phase [MWh], Energy used in data transfer (optimising packets and volume of data) [MWh], Data efficieny in terms of storage [TB] 
         & \begin{itemize} 
                 \item Energy efficient network operation 
                 \item Energy efficient devices
                 \item Energy efficient AI/ML training and inference
             \end{itemize} \\ \hline
         \vspace{0.5cm}  Life expectancy of robots [unit of time], Number of virtualized functionalities 
      & \begin{itemize} 
            \item Modularization
            \item Virtualization
            \item Softwarization
            \item Compute offloading
            \item Resilience
            \item Predictable low-latency
        \end{itemize} \\ \hline
      \vspace{0.3cm} Number of injuries at work
        & \begin{itemize} 
            \item Ubiquitous connectivity
            \item Resilience
            \item Explainability
        \end{itemize} \\ \hline
      \vspace{1cm}  Downtimes [hours], Non-conformance cost [\euro]
      & \begin{itemize}
            \item Ubiquitous connecitivity
            \item Low latency
            \item AI-native
            \item Reliability
            \item Resiliency
            \item Service exposure
            \item Ensuring SLAs
        \end{itemize} \\ \hline 
   \vspace{0.1cm} Average price of cobots [\euro], Initial investment required for setting up Collaborative Robots [\euro], OPEX from equipment, maintenance, operation, service, training, etc. of CMR [\euro], Running CAPEX [\euro]
& \begin{itemize} 
        \item Low CAPEX
        \item Closed-loop control
        \item Reliability
        \item Monitoring and telemetry
        \item Low energy/energy neutral
        \item Zero-touch 
        \item Service exposure
    \end{itemize} \\ \hline
        
    \end{tabular}
   
    \label{tab:tab3}
\end{table}


\subsection{6G E2E system enabler selection for Cobots use case} \label{sec4e}
The features for the nodes and edges of the full KG, which was obtained from the metadata collected in \cite{Jain2025} and shown in Fig. \ref{fig:fullKG}, are generated based on the transformed meta-data, as discussed in Section \ref{sec4}-\ref{sec4d}. Note that, the full KG does not include the KVI information. From Fig. \ref{fig:fullKG}, it can be observed that the full KG for the cobots use case presents a complex landscape for the enabler selection process. 


\begin{figure*}[!htb]
    \centering
    \includegraphics[width=2.1\columnwidth]{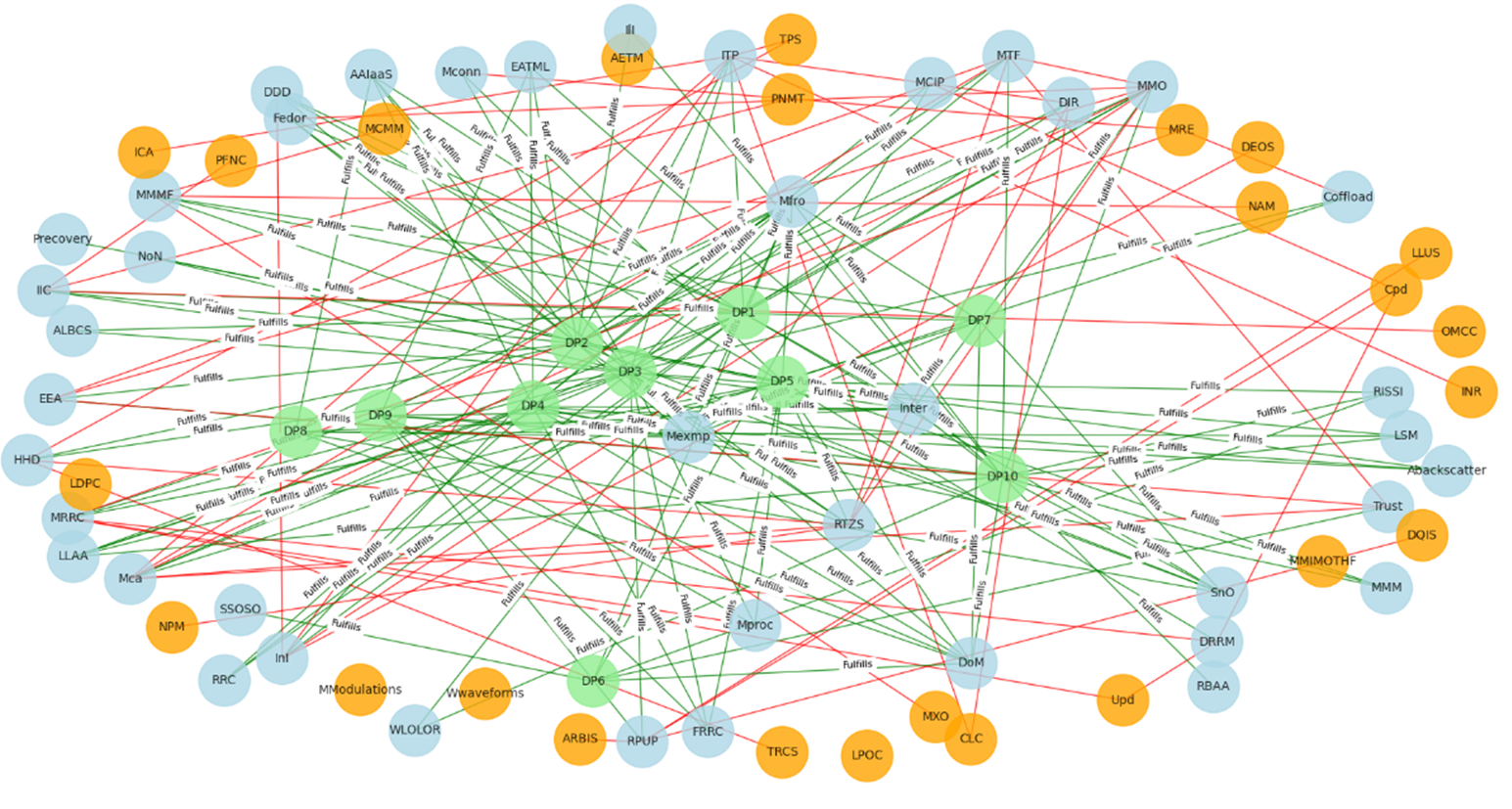}
    \caption{Pruned knowledge graph based on maturity and criticality towards migration of enablers}
    \label{fig:PrunedKMatcric}
\end{figure*}

\begin{figure}[htb]
    \centering
    \includegraphics[width=0.52\textwidth]{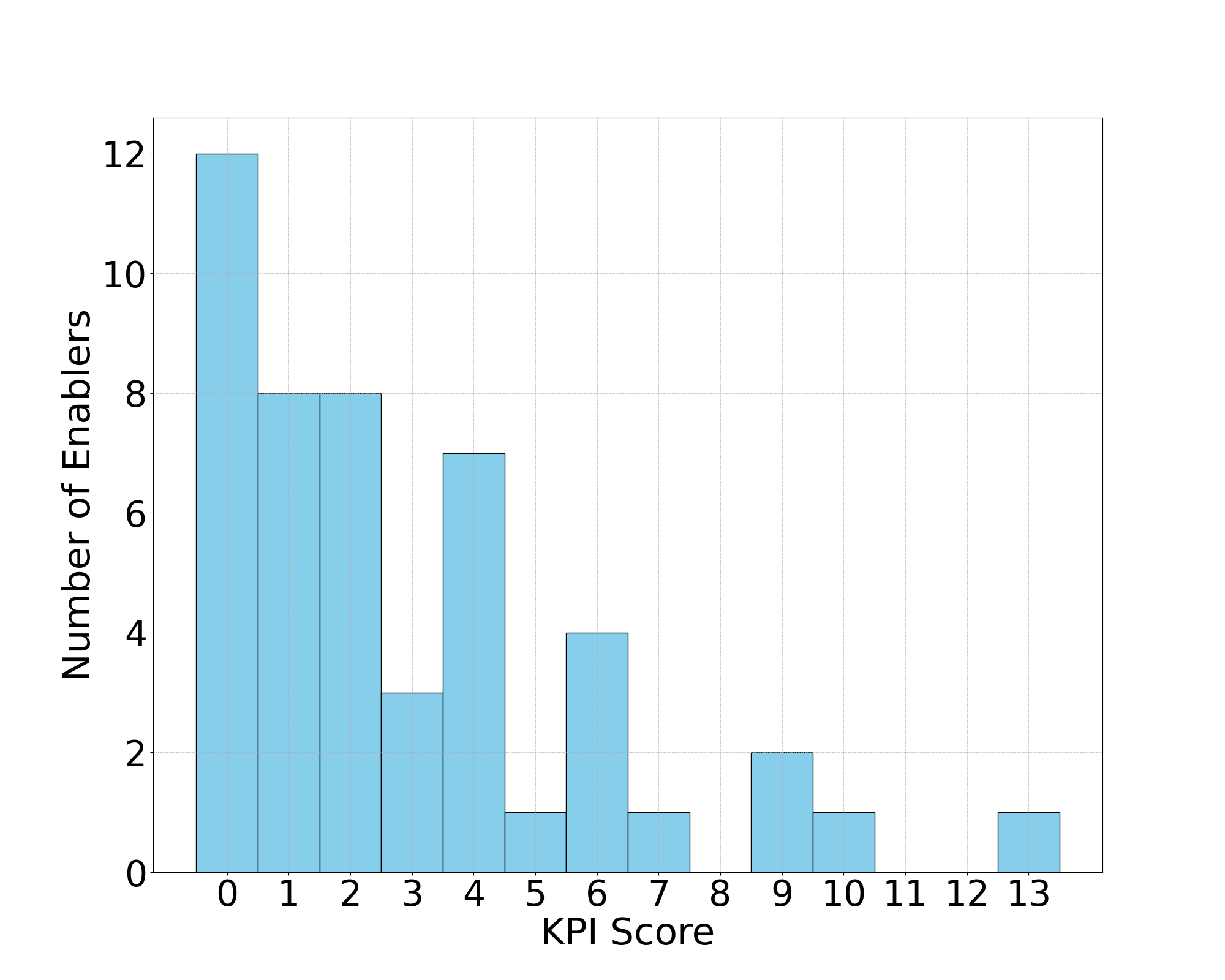}
    \caption{KPI impact analysis for each enabler in the pruned KG}
    \label{fig:KPImpact}
\end{figure}

\begin{figure*}[!htb]
    \centering
    \includegraphics[width=2.1\columnwidth]{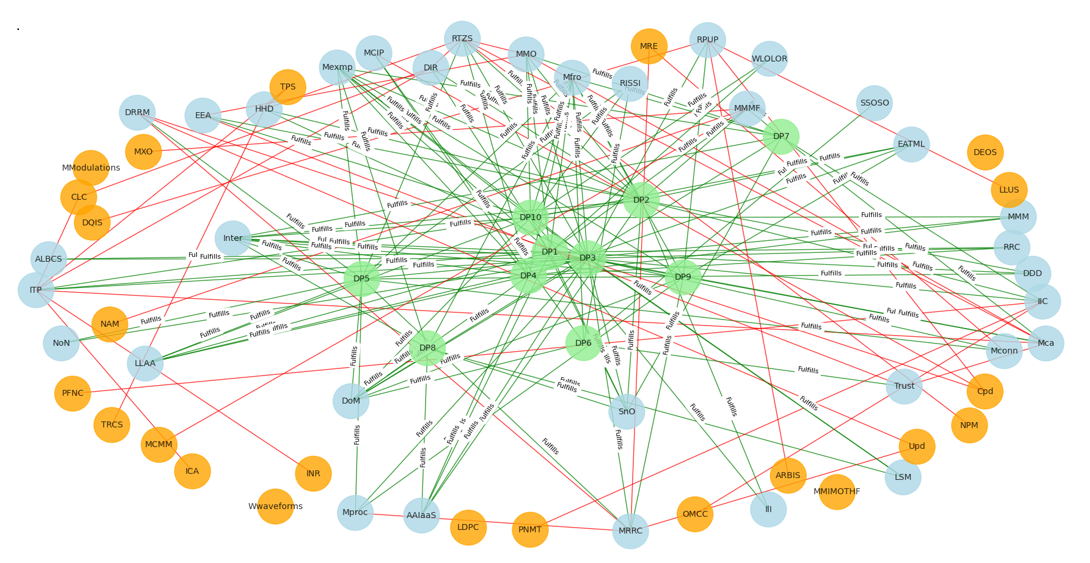}
    \caption{KPI impact based pruned KG}
    \label{fig:prunedKPIKG}
\end{figure*}

To perform the enabler selection in this complex landscape, a first pruning step on the full KG is carried out based on the maturity and criticality towards migration. The pruning is performed by thresholding the maturity feature to at least TRL 2, i.e., all enablers with at least TRL 2 are retained. Additionally, all enablers that are critical towards migration are also preserved. The KG after this pruning step is shown in Fig. \ref{fig:PrunedKMatcric}. Subsequently, an analysis of the KPI scores for each of the selected enablers, excluding the dependencies, within the pruned KG is performed. Specifically, the KPI scores for each enabler are computed by summing up the feature values of the positive, neutral and negative impact of an enabler towards achieving the use case KPIs listed in Table \ref{tab:Tab1}. 
Fig. \ref{fig:KPImpact} demonstrates a frequency analysis of the KPI scores of all the enablers within the pruned KG. From Fig. \ref{fig:KPImpact}, it can be observed that 12 enablers have a net zero KPI impact on the use case KPI requirements. This means that these enablers do not contribute to the overall fulfillment of the use case KPI requirements. Hence, as part of the graph pruning process a threshold on the KPI score of greater than or equal to 1 is set. This means that enablers which have a positive impact of at least 1 for satisfying use case KPI requirements are selected. Subsequently, the KG obtained in Fig. \ref{fig:PrunedKMatcric} is pruned to generate a KPI-based KG (KKG), which is presented in Fig. \ref{fig:prunedKPIKG}. Note that, in addition to the enablers selected based on the KPI impact analysis, enablers that were selected in our earlier work in \cite{hexa2_D23} have also been included for the steps that follow.

The selected enablers from the KKG are then categorized into multiple enabler categories based on their functionality. \textcolor{black}{This is done so that functionality similar enablers are handled consistently during subsequent KVI-based pruning.} For example, the \textit{Intelligent radio interface} enabler category will consist of enablers such as the AI transmitter, AI receiver, etc. Similarly, the enabler category of \textit{Security and privacy controls} will consist of enablers such as Level of Trust Assessment Function (LoTAF). Concretely, $23$ enabler categories are defined to capture all the selected enablers. We refer the reader to Annex-C of \cite{hexa_D25}, for a detailed list of these enabler categories alongside the selected enablers enlisted within them. 

Subsequently, a novel mapping of the enabler categories and the enablers within those categories to the technical requirements is provided. This mapping allows a direct mapping of the enablers to the KVIs related to those technical requirements. For the sake of brevity, in this article the mapping process for the \textit{AI-native air interface}, \textit{6G-based sensing}, \textit{Flexible Topologies}, \textit{LoTAF}, \textit{Intent-based management} and \textit{Integration Fabric} enablers has been presented. Table \ref{tab:catmap} presents the mapping for the aforementioned enablers to the KVIs. From the approximately $80$ enablers selected out of more than $100$ enablers being studied in Hexa-X-II and other EU projects, the aforementioned enablers were selected for discussion as they are representative of each layer of the 6G E2E system blueprint \cite{hexa2_D23}. We refer the reader to Annex-C of \cite{hexa_D25} for the mapping of the complete list of enabler categories and the technological enablers within them, to the KVIs. 

\begin{table*}[]
    \centering
    \caption{Mapping of Enablers to Enabler KVIs}
    \begin{tabular}{|p{1.5cm}|p{3.5cm}|p{4.5cm}|p{6.5cm}|} \hline
         \textbf{Enabler}& \textbf{Enabler Category} & \textbf{Technical Requirements Addressed} & \hspace{2.3cm}\textbf{KVIs Addressed}  \\ \hline
         AI-native air interface & Intelligent radio interface & \begin{itemize} 
             \item AI-native
             \item Reliability to varying channel conditions
             \item Improved data rates and spectral resource efficiency
        \end{itemize} & \begin{itemize} 
             \item Lower downtimes and hence, improved \textbf{efficiency} 
             \item Reduced \textbf{costs} due to improved coverage, throughput and latency
             \item Improved \textbf{Trustworthiness} in network performance
         \end{itemize} \\ \hline
        6G-based sensing & ISAC physical layer and beyond communications services & \begin{itemize}
            \item Reliability to do improved sensing and localization
            \item Signaling and protocols for ISAC
        \end{itemize} & \begin{itemize} 
            \item Lower downtimes and hence, improved \textbf{efficiency} 
            \item Industry specific production KPIs demonstrating improved  \textbf{Productivity}
        \end{itemize} \\ \hline
        Flexible Topologies & Flexible radio interface and protocol & \begin{itemize}
            \item Modularization
            \item Energy efficient network operation
            \item Reliability and resilience to adverse channel conditions and scenarios
            \item Low latency
        \end{itemize} & \begin{itemize} 
            \item Reduced \textbf{total energy usage}
            \item Lower downtimes and hence, improved \textbf{efficiency} 
            \item Industry specific production KPIs demonstrating improved  \textbf{Productivity}
        \end{itemize} \\ \hline
        LoTAF & Security and Privacy controls & \begin{itemize} 
            \item Trust and Security
        \end{itemize} & \begin{itemize}
            \item \textbf{Trustworthiness} due to provision of trust and security mechanism
        \end{itemize} \\ \hline
        Intent-based management & Intent-based management & \begin{itemize}
            \item Zero-touch control and automation 
            \item Virtualization, Modularization and Softwarization of network functions
        \end{itemize} &  \begin{itemize}
            \item Improved productivity 
            \item Reduced \textbf{material usage/waste} through virtualization and modularization of network functions/components
        \end{itemize}\\ \hline
        Integration fabric & Service exposure, Data framework, Programmable and autonomous networks, Security and privacy controls and Cloud-continuum integration  & \begin{itemize} 
            \item Modularization, Softwarization and Virtualization of network functions
            \item Monitoring and Telemetry
            \item Exposure of management capabilities via a service exposure framework
            \item Trust and Security 
        \end{itemize} & \begin{itemize} 
            \item \textbf{Trustworthiness} due to multiple levels of trust functions
            \item Improved productivity 
            \item Reduced \textbf{material usage/waste} through virtualization and modularization of network functions/components
            \item Reduced \textbf{costs} in OPEX due to intuitive inputs and feedback for IT/OT teams in industries, as well as improved revenues due to exposure framework
            \item Improved \textbf{safety} in work place due to continuous monitoring and availability of data
        \end{itemize} \\ \hline
    \end{tabular}
    
    \label{tab:catmap}
\end{table*}

\begin{figure}[!htb]
    \centering
    \includegraphics[width=\columnwidth]{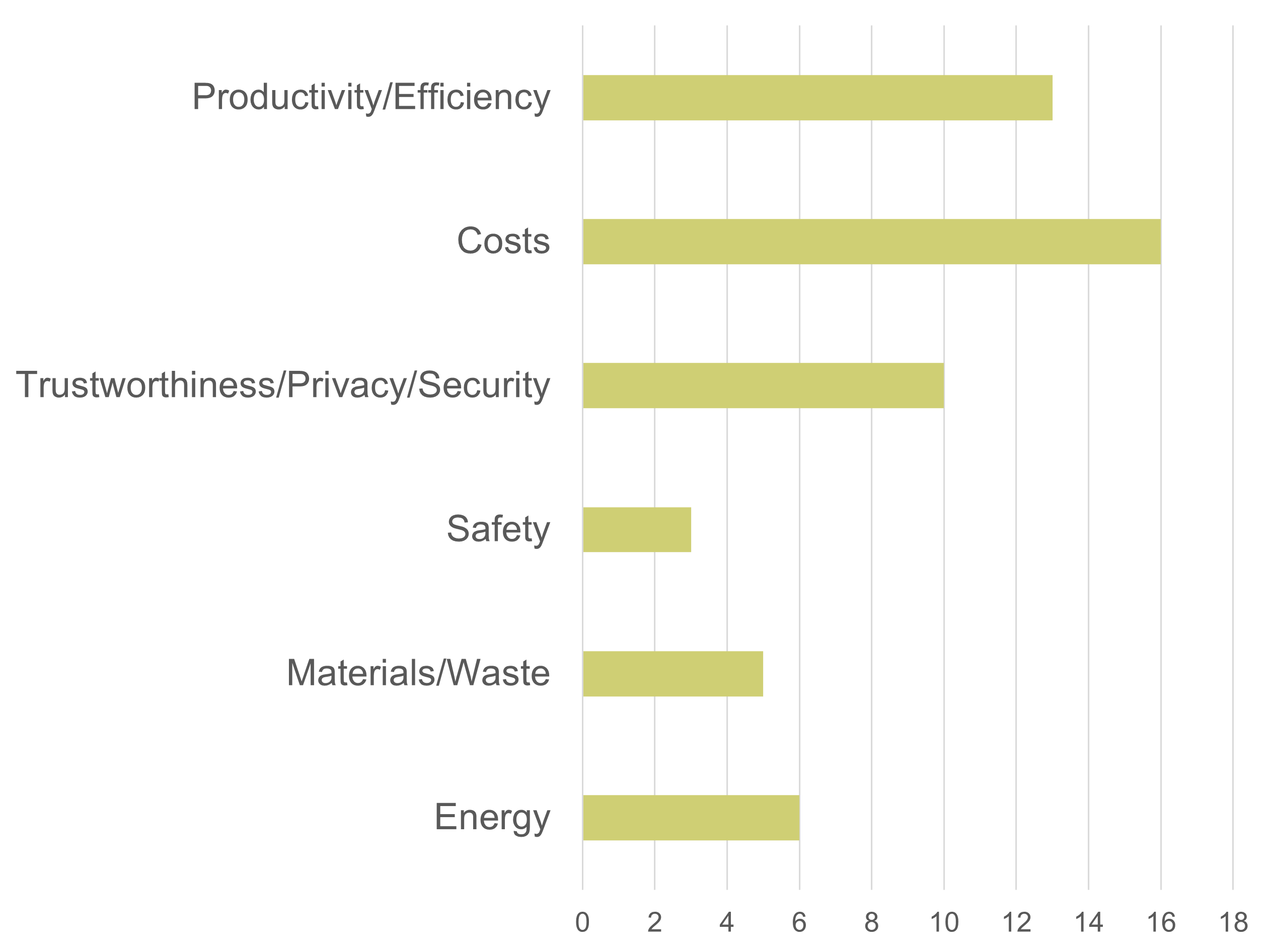}
    \caption{Number of Enabler categories per KVI category}
    \label{fig:KVIanalysis}
\end{figure}

From Table \ref{tab:catmap} it can be observed that each of the enabler maps to a certain set of technical requirements. These are then utilized, based on the mapping shown in Table \ref{tab:tab3}, to map the enablers to the KVIs they address. Note that, some of the enablers discussed map to multiple enabler categories as they are a combination of multiple enablers from these different enabler categories. For example, two of the multiple enablers that the integration fabric is composed of are the monitoring and telemetry framework, which is part of the Data framework category \cite{hexa_D25}, and Management and capabilities exposure, which is part of the Service exposure category \cite{hexa_D25}. Next, it can be observed that the enablers collectively satisfy KVIs related to energy efficiency, material efficiency, costs, trustworthiness, safety, increased productivity as well as improved revenues. When all $23$ enabler categories are subjected to a similar analysis and then the number of enabler categories that satisfy the six broad classes of KVIs, which have been introduced in Table \ref{tab:tab2}, are computed, we obtain Fig. \ref{fig:KVIanalysis}.  

\begin{table}[!htb]
\caption{\textcolor{black}{Comparison of proposed approach with closest baselines for the COBOTs use case.}}
\label{tab:baseline_kpi_kvi}
\centering
\begin{tabular}{|p{0.12\linewidth}|p{0.20\linewidth}|p{0.15\linewidth}|p{0.33\linewidth}|}
\hline
\textbf{\textcolor{black}{Method}} & \textbf{\textcolor{black}{Selection Methodology}} & \textbf{\textcolor{black}{\# Enablers}} & \textbf{\textcolor{black}{Key implication (COBOTs)}}\\
\hline
\textcolor{black}{KPI-only (baseline)} &
\textcolor{black}{KPI-threshold pruning (KPI score $\geq 1$) + pragmatic consideration} &
\textcolor{black}{$70$} &
\textcolor{black}{KPI-feasible candidate set, but may reduce KVI coverage (notably safety) by excluding low-KPI enablers with unique KVI contribution.}\\
\hline
\textcolor{black}{Proposed (KPI+KVI, KG-based)} &
\textcolor{black}{KPI pruning + KVI analysis + pragmatic consideration} &
\textcolor{black}{$82$} &
\textcolor{black}{Preserves KPI feasibility and makes KVI coverage explicit; closes KVI gaps via targeted reintroduction (e.g., Pcell recovery for safety).}\\
\hline
\textcolor{black}{Blueprint-level enabler list \cite{hexa2_D23}} &
\textcolor{black}{Curated list after pragmatic considerations \cite{hexa2_D23}} &
\textcolor{black}{$45$} &
\textcolor{black}{Provides a blueprint/architecture-level selection, but does not explicitly operationalize KPI/KVI scoring and coverage analysis as in the proposed workflow (granularity differs from the technology-enabler pool used here).}\\
\hline
\end{tabular}
\end{table}

The KVI analysis in Fig. \ref{fig:KVIanalysis} demonstrates that the enablers determined by the KG method for the 6G E2E system improve the industrial productivity and efficiency. This is achieved via low downtimes and automation, which is also visible in the enabler to KVI mapping in Table \ref{tab:catmap}. These enablers also provide improved monetization and reduction in OPEX. Additionally, adequate trust, security and privacy based enablers have been selected by the KG method, thus provisioning satisfactory level of trustworthiness, privacy and security. Next, the designed 6G E2E system also consists of satisfactory number of enablers to reduce energy usage and material usage\slash waste. One of the main drawbacks within the set of selected enablers is the lack of enablers to further improve workplace safety. Hence, as part of the pragmatic consideration step the \textit{Pcell recovery} enabler, which was discarded during the KPI-based pruning step due to low KPI fulfillment criteria, is reintroduced. It provisions enhanced safety characteristics, as it will enable service continuity and connectivity for COBOTs even if the primary serving cell becomes blocked. \textcolor{black}{This is done by moving the connection to a secondary cell. Consequently, it also improves the overall KVI coverage offered by the set of selected enablers.} This enabler is categorized as part of the  \textit{Flexible radio interface and protocol} enabler category. 

The final list of the 82 selected enablers can be found in Annex-C of \cite{hexa_D25}. \textcolor{black}{This set of enablers represents a design-time candidate pool for 6G E2E system design. Even at design time, pruning provides tangible benefits by reducing the integration and verification burden of the E2E system (e.g., fewer candidate interfaces, fewer cross-domain dependencies to validate, and reduced orchestration complexity), which translates into lower engineering effort and cost. The enablers removed during pruning are primarily redundant for the COBOTs use case because they either exhibit low overall KPI and KVI impact under the selected thresholds or provide functionality that is already covered by higher-impact retained enablers. }


\textcolor{black}{Table~\ref{tab:baseline_kpi_kvi} summarizes how the proposed KG-based approach improves over the closest baselines, i.e., a KPI-only selection obtained via KPI-threshold pruning and a blueprint-level reference enabler list from  \cite{hexa2_D23}. Relative to KPI-only pruning, the proposed approach preserves KPI feasibility while making KVI coverage explicit and allowing targeted reintroduction of enablers to close gaps (e.g., safety). Relative to \cite{hexa2_D23} blueprint-level selection, our method provides a requirement-driven and auditable procedure that explicitly considers both  KPIs and KVIs through scoring, pruning, and coverage analysis. This proposed approach is consequently able to provision justification as to why enablers are retained or reintroduced for any given use case.}

With this background, the KG method can be utilized similarly to design the 6G E2E system for other use cases. \textcolor{black}{For example, immersive XR can be modeled through stringent throughput/latency requirements, while remote healthcare can be modeled through high reliability and trust requirements, and the same KG pruning and KVI analysis workflow can be applied by updating the requirement inputs accordingly. In general, it is envisioned that a large proportion of the enablers will be the same for different use cases. Enablers} such as AI-native air interface or LoTAF , which provision high throughput, low latency and increased trust, are necessary and beneficial for any use case. Hence, such enablers will form a common set of enablers that will transcend multiple use cases. On the other hand, enablers such as 6G-based sensing will be of utility in certain use cases, such as the one under discussion in this section, while being of reduced importance for use cases such as mobile broadband services. \textcolor{black}{However, deriving a detailed common set of enablers across multiple use cases is beyond the scope of this paper and is left as future work}.

\textcolor{black}{Furthermore, in the proposed methodology, use case KPIs act as primary feasibility constraints, ensuring that use case KPI requirements are not violated during pruning. Sustainability-oriented KVIs are then used to prioritize among feasible enabler sets and to improve overall system value (e.g., trustworthiness, productivity, OPEX reduction, material usage). This naturally introduces tradeoffs. For instance, an enabler may offer strong KVI coverage (e.g., safety/trust) while providing a weaker KPI contribution, motivating the pragmatic reintroduction step described earlier. Therefore, KVIs do not replace hard-performance goals. Instead, they complement KPI-driven pruning by guiding selection among KPI-feasible alternatives.}


\section{SELECTED ENABLERS PERFORMANCE AND SUSTAINABILITY ANALYSIS THROUGH PROOF OF CONCEPT-BASED VALIDATION} \label{sec5}

This section highlights the performance and sustainability validation of a subset of the selected 6G enablers, listed in Table \ref{tab:catmap}, through PoC experimentation conducted across Hexa-X-II. These enablers span multiple network layers and architectural domains—including AI-native air interfaces, joint communication and sensing, flexible topologies, trust-driven orchestration, intent-based management, and integration fabrics. \textcolor{black}{This paper differs from prior work because} each enabler has been quantitatively assessed against system-level KPIs and KVIs, with experimental setups capturing throughput gains, latency profiles, energy savings, CPU overhead, and system availability. Their inclusion in the 6G E2E system is further supported through knowledge graph-based selection, ensuring strong alignment with core 6G design principles and sustainability ambitions across environmental, economic, and social dimensions. \textcolor{black}{Note that, the PoC-based validation focuses on enablers with demonstrated maturity (TRL 4 and above) to provide experimental evidence that the KG-selected enablers deliver the intended KPI/KVI effects in realistic environments.}

The following subsections provide empirical validation of these enablers using realistic testbeds and simulations. Key metrics include throughput gains, latency distributions, CPU load, energy efficiency, and orchestration responsiveness. These results not only confirm the technical viability of each enabler but also reinforce the relevance of the KG-based selection approach in guiding architectural decisions towards a performant and sustainable 6G system.

\subsection{AI-native air interface}\label{sec:aiairintf}

One of the emerging concepts in the realm of 6G is a so-called AI-native air interface. This refers to a network design where AI is considered in all layers of the radio system. Focusing on the physical layer, one approach for implementing such an AI-native air interface is to utilize a partially learned transmit signal with an AI-based receiver. This means that the waveform is implicitly learned such that it suits the purposes of the AI receiver.

\begin{figure}[h!]
    \centering
    \includegraphics[width=\columnwidth,trim=0 400 650 0, clip]{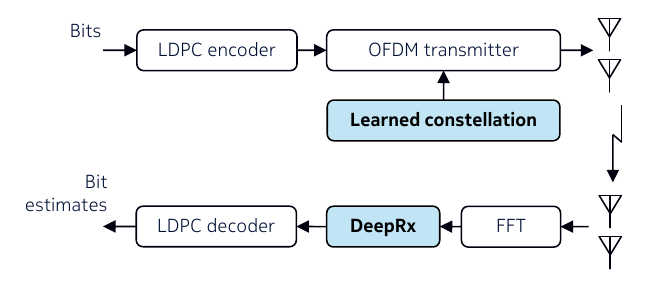}
    \caption{A simplified depiction of the AI-native air interface concept, where learned constellations are used jointly with AI-based DeepRx receivers.}
    \label{fig:aiai_concept}
\end{figure}

Such a concept has been developed earlier by utilizing a learned constellation with a DeepRx-type receiver, and learning the constellation parameters and the weights of the receiver algorithm jointly \cite{Korpi2023}. This scheme is illustrated in Fig.~\ref{fig:aiai_concept}. In particular, the approach involves learning the transmitter constellation and receiver algorithm (DeepRx) to achieve a system which can communicate without any channel estimation pilots. This will result in throughput gains via the reduced overhead. Note that the system is still based on the widely-used orthogonal frequency-division multiplexing (OFDM) waveform, wherein only the constellation diagram is modified.

The performance gain of such an approach has been evaluated both with simulations \cite{Korpi2023} and with measurements \cite{hexa2_D45}. The simulations were performed within a multiple input and multiple input (MIMO) link with two spatially multiplexed streams, and throughput gains in the order of 15--20\% were observed. The over-the-air measurements were carried out with a single input and single output (SISO) system in an indoor scenario. There, average throughput gain of 19.2\% was observed. In both cases, the throughput gain is mostly attributed to the reduced overhead, since the AI-native air interface scheme can transmit data without channel estimation pilots, owing to the learned constellation.

Essentially, such an AI-native air interface improves the spectral efficiency of the wireless network. The increased efficiency can be used in various ways to improve the system performance, e.g., to improve throughput or increase energy efficiency. As long as the AI inference can be done with high enough energy efficiency, the latter approach contributes directly to the sustainability targets of 6G networks, which has been one of the main ambitions of the Hexa-X-II project. Moreover, this validates the selection of the AI-native air interface enabler by the KG method, as it has demonstrated itself to be at TRL 5-6 with extremely positive simulation and measurement-based results towards improvement of throughput, reduction in energy consumption as well as trust in network performance. Specifically, these results clearly elaborate the high KPI and relevant KVI contribution of the AI-native air interface towards the use case requirements. Additionally, it also demonstrates alignment with Hexa-X-II design principles, i.e., flexibility to different network scenarios, resilience and sustainability \cite{hexa2_D21}.  

\subsection{6G-based sensing} \label{sec:6Gbasedsens}
Sensing through telecommunications infrastructure is emerging as a pivotal enabler for future 6G systems. It offers telecom vendors promising new business opportunities, while enabling innovative services for end-users. These include collision avoidance for automotive applications, enhanced industrial autonomy such as automated warehousing, healthcare services like fall detection in elderly care, and broader smart city solutions. The integration of sensing into telecom networks transforms them into dual-purpose systems, significantly increasing their societal and commercial value.

\begin{figure}[h!]
    \centering
    \includegraphics[width=\columnwidth]{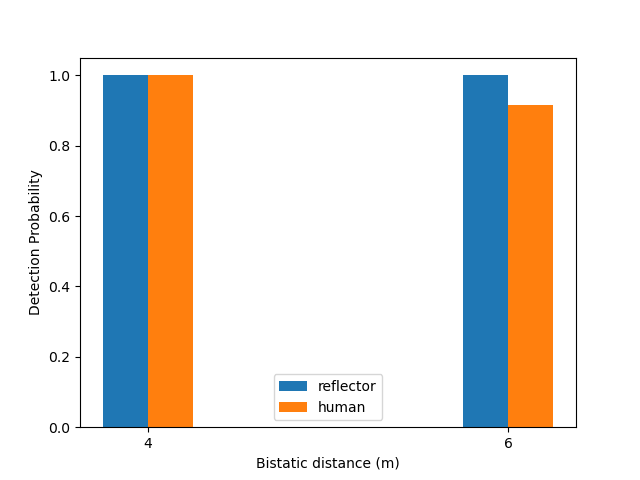}
    \caption{Percentiles for detecting a person passing in the corridor in an indoor office space for different bi-static distances and within half a meter accuracy}
    \label{fig:percentile}
\end{figure}

Within the Hexa-X-II project, a hardware-based enabler has been developed and demonstrated for integrated communication and sensing using adapted 5G NR signals in a bi-static radar setup. This work confirms that CP-OFDM-based telecom signals, combined with precise beamforming, are highly effective for sensing. Measurement campaigns were conducted using unlicensed mmWave spectrum with bandwidths up to 800 MHz in various environments, including indoor office settings and outdoor vehicle detection scenarios. In the indoor setting, 95th percentile detection performance with sub-meter accuracy was achieved across different bi-static distances, as shown in the accompanying Fig.~\ref{fig:percentile}. Additionally, the same telecom-based sensing platform was employed with convolutional neural networks (CNN) for gesture recognition, achieving classification accuracies above 93\%.

These results serve as a successful proof-of-concept for integrated sensing and communication in the Hexa-X-II project. They substantiate the feasibility and potential of 6G sensing, despite remaining challenges such as the need for tight synchronization across distributed nodes. Additionally, these results also validate the selection of the 6G-based sensing enabler by the KG method given that it has demonstrated a TRL of 4-5 with a hardware-based demonstration and high contribution towards the use case KPIs. Furthermore, the results show its potential in improving efficiency and productivity in factory floors through accurate sensing and localization, thus also demonstrating relevant KVI contribution. It also adheres to 6G system design principles, i.e., resilience, automation, optimization and sustainability \cite{hexa2_D21}.  

\subsection{Flexible topologies}\label{sec:flextop}
Trustworthy flexible topologies represent a key innovation in 6G architectures, addressing the need for dynamic, resilient, and secure network deployments in scenarios where traditional infrastructure is limited or compromised. In Hexa-X-II, this enabler is demonstrated through the integration of Unmanned Aerial Vehicle (UAV)-assisted mesh networking, orchestrated through AI-driven controllers and evaluated using trust metrics spanning QoS, security, and energy awareness dimensions. 

\begin{figure*}[]
    \centering
    \includegraphics[width=1.5\columnwidth]{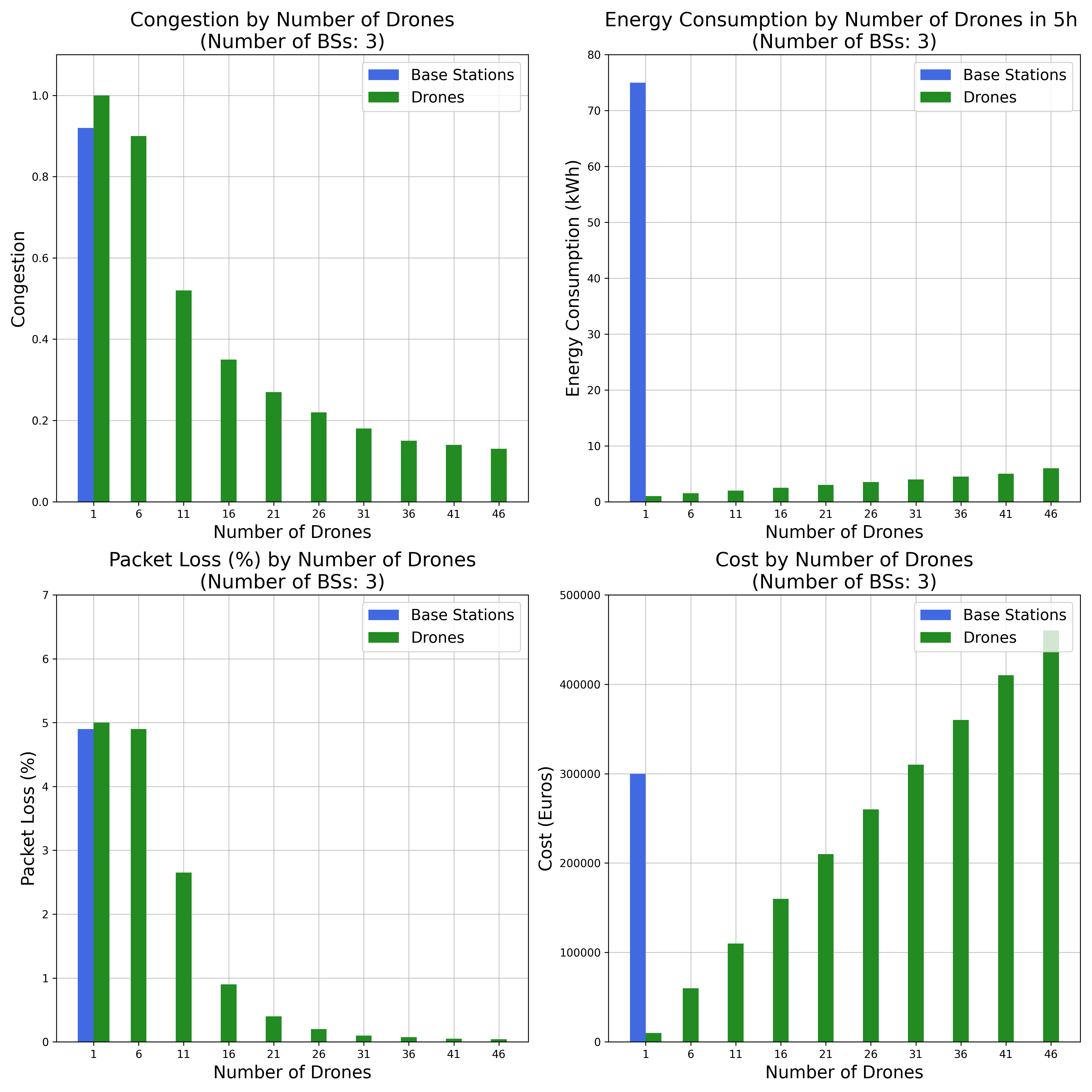}
    \caption{Evaluation of the trustworthy flexible topology enabler under increasing numbers of flexible nodes.}
    \label{fig:flextop}
\end{figure*}

At the core of the framework lies the Ad-hoc Network Controller (ANC), which coordinates UAVs operating as mobile access points (APs). These UAVs augment terrestrial nodes by adapting their positions based on real-time traffic predictions, environmental conditions, and trustworthiness indicators. The Trust Evaluation Function (TEF), extended with UAV-specific metrics (e.g., flight stability, energy levels, link integrity), ensures that only nodes meeting required thresholds contribute to the network.

The Flexible Mesh Selection Function (FMSF) applies a composite trustworthiness index, computed using both functional (e.g., latency, throughput) and non-functional (e.g., energy efficiency, operational cost) parameters.

Evaluations based on both simulations and integrated test environments yielded the following improvements (see Fig.~\ref{fig:flextop}):
\begin{itemize}
    \item Significant reduction in network congestion as the number of UAVs increases, enabling improved load balancing and system scalability,
    \item More than 90\% \% lower energy consumption for UAV-based communication compared to traditional base stations, supporting energy-efficient operation,
    \item Significant decrease in packet loss with increasing UAV density—approaching zero with more than 20 drones—highlighting improved reliability and service continuity,
    \item Linear growth in deployment cost with drone count, enabling scalable and temporary coverage expansion while avoiding the high fixed cost of dense infrastructure.
\end{itemize}

This enabler supports Environmental sustainability by minimizing energy waste and supporting energy harvesting Economic sustainability by reducing reliance on fixed terrestrial infrastructure, as well as Social sustainability by enabling secure and trusted connectivity in public safety and remote operations, leading to increased inclusion. In the KG-method pruning, this enabler was retained based on demonstrated TRL 4--5 maturity through realistic simulations, strong KVI alignment, including trust, resilience, energy efficiency and reduced CAPEX, as well as strong alignment with Hexa-X-II design principles, i.e., decentralization, autonomy, trust-awareness, and closed-loop operation.

\subsection{LoTAF}\label{sec:lotaf}

As part of the Hexa-X-II Security, Privacy, and Resilience (SPR) controls \cite{hexa_D25}, the Level of Trust Assessment Function (LoTAF) addresses the sustainability objectives by promoting efficient network operation and reducing the cost of a continuous monitoring service to ensure the fulfillment of a Trust Level Agreement (TLA). LoTAF brings a trustworthy and efficient coordination of distributed resources, since it deals with trust as a new type of intent to manage and orchestrate networks. When it comes to the sustainability pillars (see Table \ref{tab:tab2}), LoTAF contributes to the environmental aspects by recommending stable and more reliable network resources for stakeholders and by leveraging knowledge graphs to minimize information duplication and guarantee data efficiency in terms of storage and complex semantic inquiries. With respect to the economical aspects, LoTAF contributes by reducing downtime via assigning tasks to highly trusted resources, thus reducing the probability of failures and errors. In addition, it also reduces OPEX via automation and zero-touch management, enhances service assurance, and improves dynamic service migration by adaptive resource management when there are deviations in the TLA. 

\begin{figure}[h!]
    \centering
    \includegraphics[width=\columnwidth]{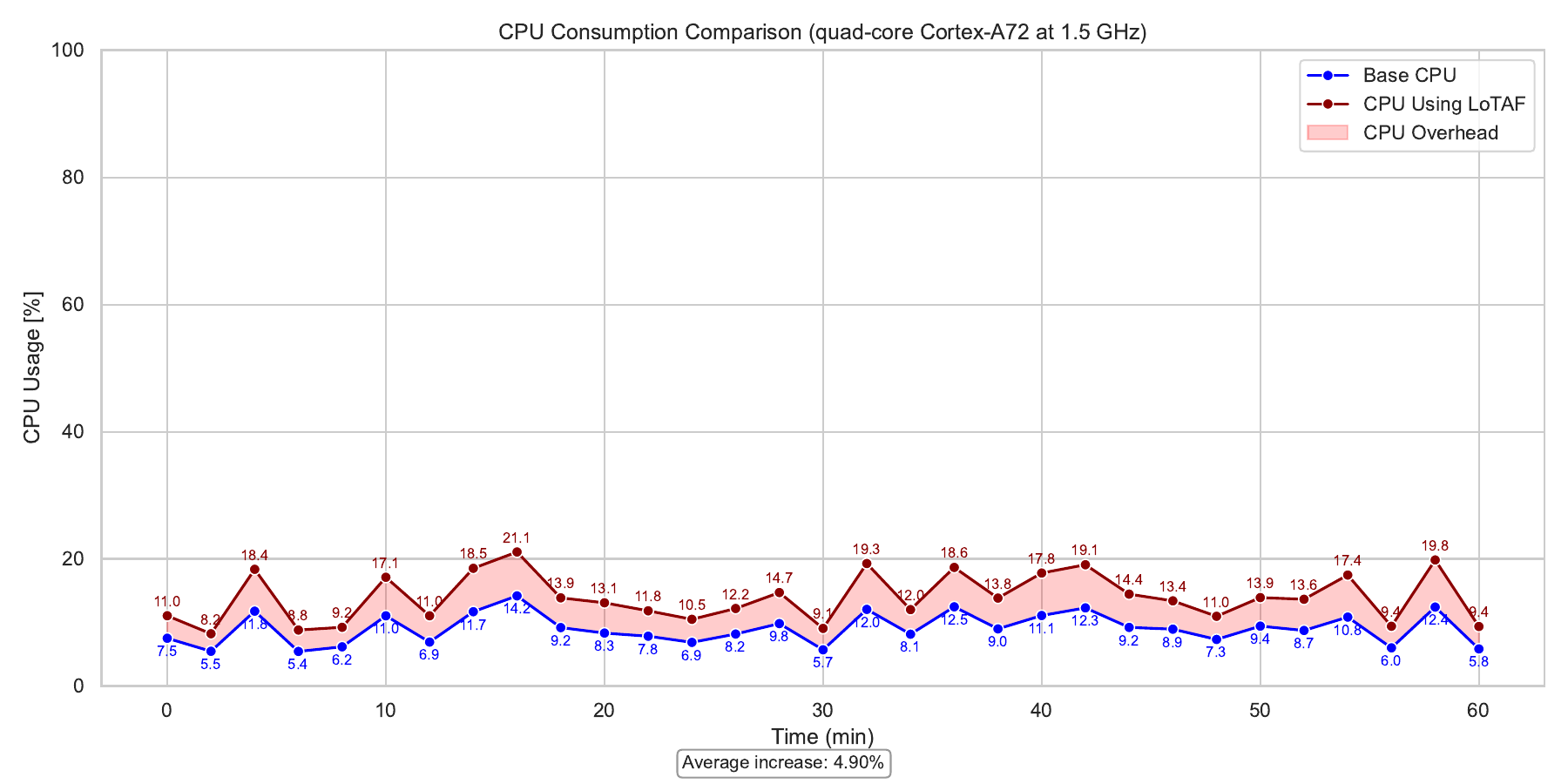}
    \caption{CPU overhead of using LoTAF on an idle device}
    \label{fig:lotaf_cpu}
\end{figure}

This selected enabler has been tested in the context of an inventory warehouse management PoC powered by cobots \cite{hexa_D26}. The experimental results showcase that LoTAF maintains high stability with an average latency of 1.1069 seconds for its complete workflow execution \cite{hexa_D26}. Such a latency is slightly increased when LoTAF is integrated with the Trust Evaluation Function (TEF) and the Integration Fabric, reaching an E2E latency from publishing trust management request until its final trust computation of 2.7369 seconds. Introducing less than 3 seconds of overhead for complete workflow execution underline the feasibility of contemplating trust as a new intent to manage and orchestrate 6G networks in a reasonable time. Likewise, LoTAF demonstrates a negligible CPU overhead over an idle device like a Raspberry Pi 4 (RPi4) with the ROS2 system running (see Figure \ref{fig:lotaf_cpu}). After setting the monitoring agents to continuously gather health parameters on the device, CPU usage only increased by 4.90\% over the benchmark, that is, use of the device with a minimal operating system and no additional applications running. 

With this background, it can be stated that LoTAF is at TRL 4-5, wherein it provisions improved trust within the 6G system through TLAs. Moreover, it does not increase the latency by a significant amount and also keeps the CPU utilization at a satisfactory level. As a result, LoTAF provisions strong KPI and KVI alignment. It also satisfies the Hexa-X-II design principles, i.e., trust-awareness and closed-loop operations. Hence, the selection of LoTAF reinforces the effectiveness of the KG method for enabler selection and 6G E2E system design. 


\subsection{Intent-based Management}\label{sec:ibm}
The selected intent-based management (IBM) solution, called intent-based network intention management entity (IBN-IME), was validated in independent experimental environments \cite{alemany_ecoc23}\cite{adanza_netsoft24}. From a sustainability perspective, IBM-IME supports all three sustainability pillars in Table \ref{tab:tab2} as follows
\begin{itemize}
    \item economically, by reducing OPEX through automation, zero-touch operations, and efficient service orchestration. These benefits are reflected in Table \ref{tab:tab2} "Enabler KVIs" such as reduced "OPEX from equipment, maintenance, operation, service, training" and improved "downtimes".
    \item environmentally, by reducing unnecessary resource utilization, contributing indirectly to lower "energy consumption per process" and improved "data efficiency in storage and transfer".
    \item socially, by enhancing resilience and stability of network operations that contribute to a safer operational environment, which aligns with KVIs such as reduction in "non-conformance cost" and improved service availability, indirectly contributing to safer work environments. 
\end{itemize}

\begin{figure}[!htb]
    \centering
    \includegraphics[width=1\columnwidth]{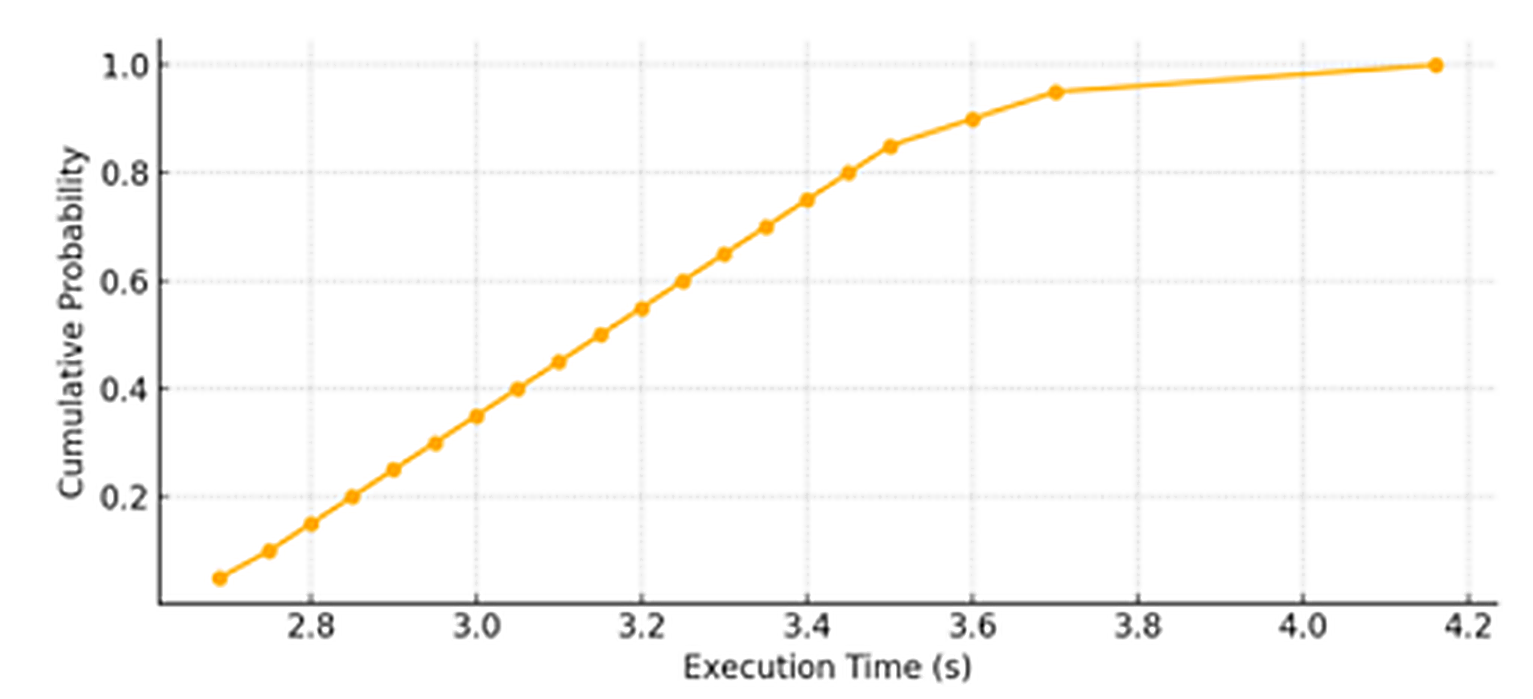}
    \caption{Total execution time CDF of the intent-based management requests.}
    \label{fig:ibn_results}
\end{figure}

The selected enabler, through the IBN-IME, has been used in two of the designed Hexa-X-II Proof of Concepts (PoCs) \cite{hexa_D24}\cite{hexa_D26}. In both the PoCs, the enabler was very stable in terms of execution time, which allowed us to obtain the Cumulative Distribution Function (CDF) of the total processing time as illustrated in Figure \ref{fig:ibn_results}. Within this processing time, the executed actions included were: intent interpretation (i.e., from human-format to 3GPP-format data objects), database storage, feasibility assessment (i.e., validate there are the resources to deploy the service), request generation (e.g., SDN-based requests using the ETSI TeraFlowSDN), and a message publication/distribution in the Integration Fabric used. 

The presented CDF indicates low variability with execution times ranging from 2.69 s to 4.16 s. This stability comes primarily from deterministic stages such as intent interpretation, database storage, and feasibility evaluation. In contrast, the TFS request generation and Kafka publishing stages exhibit higher variability due to their dependence on external systems. The obtained average execution time was approximately 3.10 s, when contextualized against the system's operational KPIs, this time aligns well with the intent management latency expectations for non-real-time management operations, as opposed to stringent ultra-low-latency communication for the data plane (Table \ref{tab:Tab1}).

The KG-based enabler selection framework further justifies the inclusion of IBM by evaluating its strong alignment with multiple design principles (e.g., modularization, zero-touch operation, CL control, etc.). In the KG pruning process, IBM demonstrated: 
\begin{itemize}
    \item high KPI contribution (e.g., supporting SLA adherence, improving operational efficiency, etc.)
    \item relevant KVI contribution (e.g., addressing both OPEX reduction and resilience-related KVIs)
\end{itemize}

In summary, the IBM enabler fulfills a key role in simplifying network operations and also contributes directly to the sustainability goals of the 6G E2E system.

\subsection{Integration Fabric} \label{sec:if}
\begin{figure}[!htb]
    \centering
    \includegraphics[width=1\columnwidth]{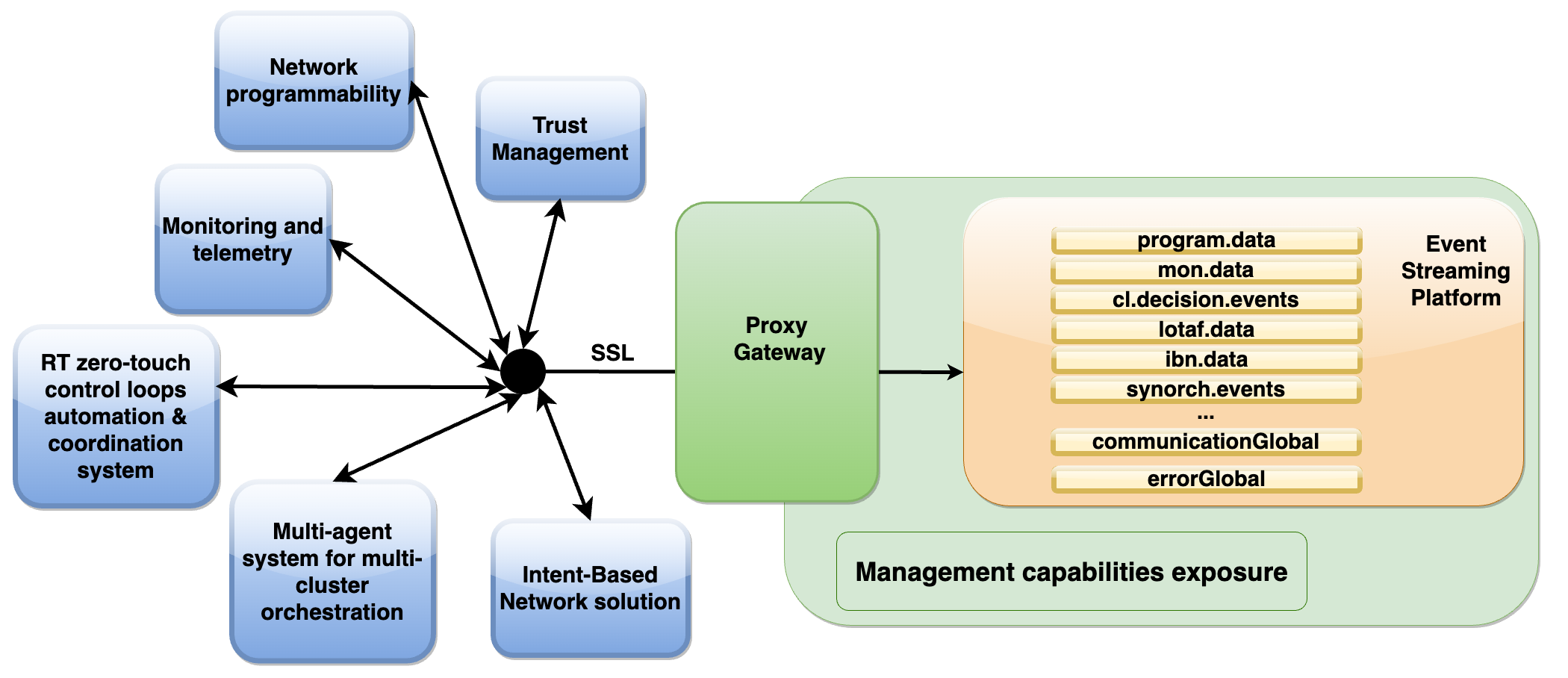}
    \caption{Integration fabric enablers interaction}
    \label{fig:if}
\end{figure}

The integration fabric, as defined in the 6G E2E system in Hexa-X-II \cite{hexa_D26}, plays a critical role in advancing sustainability objectives by enabling efficient coordination of distributed resources, enhancing service exposure, and ensuring seamless operation across multiple domains. Leveraging an event-driven, topic-based architecture with Apache Kafka as a real-time streaming data pipeline allows the integration fabric to support asynchronous, always-on communication between intent management, orchestration and observability entities \ref{fig:if} . This design transition from traditional client-server or RPC-based models to a shared, encrypted topic structure directly contributes to the sustainability pillars. Specifically, it reduces OPEX through automation and zero-touch management, improves service resilience by supporting closed-loop automation and dynamic service migration, and enhances resource and energy efficiency via adaptive resource management and virtualization. These capabilities align with KVIs such as energy consumption per process, OPEX reduction and improved service uptime, as outlined in Section \ref{sec4}-\ref{sec4b}. Experimental results demonstrate that the integration fabric maintains high stability, with average latencies between 1.66 and 3.78 seconds, 100\% system availability, and consistent throughput, introducing less than one second of overhead for complete workflow execution \cite{hexa_D26}. These metrics align with the target KPIs for management and orchestration in demanding 6G scenarios, supporting requirements for high availability, scalability, and low operational latency as specified for use cases like cooperating mobile robots, where seamless coordination and real-time responsiveness are essential.

Furthermore, the enabler’s inclusion in the system design is justified through the KG-based selection methodology. The integration fabric fulfills essential design principles such as modularization, virtualization, resilience, and zero-touch operation, and its selection is validated through KG-based pruning, ensuring it directly supports both the technical and sustainability ambitions of the 6G E2E system.

\begin{table*}[t]
\caption{\textcolor{black}{Concluding validation of KG-selected enablers (Section V.A--V.F): PoC evidence, sustainability pillars, and KVI-driven selection rationale.}}
\label{tab:secV_concluding_validation}
\centering
\renewcommand{\arraystretch}{1.15}
\textcolor{black}{%
\begin{tabular}{|p{0.15\linewidth}|p{0.30\linewidth}|p{0.10\linewidth}|p{0.32\linewidth}|}
\hline
\textbf{Enabler (Sec.)} & \textbf{PoC evidence (KPI/KVI-linked)} & \textbf{Sustainability pillars} & \textbf{Validation of KG selection (KVI-driven pruning)} \\
\hline
\textbf{AI-native air interface (V.A)} &
Throughput gain $\sim$15--20\% (simulated); avg.\ 19.2\% (OTA); reduced pilot overhead; supports energy-efficiency use of spectral gains; trust in performance &
Environmental, Economic, Social (trust) &
Retained due to strong KPI contribution (throughput/spectral efficiency) and KVI alignment (lower downtimes/efficiency, reduced cost, trustworthiness); PoC maturity TRL 5--6 confirms feasibility \\
\hline
\textbf{6G-based sensing (V.B)} &
95th-percentile sub-meter detection in ISAC PoC; gesture recognition $>$93\% (CNN); hardware demo; supports factory-floor efficiency/productivity &
Economic, Social &
Selected due to high KPI relevance for sensing/localization requirements and KVI alignment (productivity/efficiency via sensing-enabled autonomy); PoC TRL 4--5 substantiates inclusion \\
\hline
\textbf{Flexible topologies (V.C)} &
Reduced congestion with more UAVs; $>$90\% lower energy vs.\ base-station comms; packet loss approaches 0 with $>$20 drones; scalable cost trend; resilient coverage &
Environmental, Economic, Social &
Retained based on KVI alignment (reduced energy use, reduced CAPEX reliance on fixed infra, resilience/trust) and KPI evidence (reliability/continuity); PoC results validate KG pruning outcome \\
\hline
\textbf{LoTAF (V.D)} &
Workflow latency 1.1069\,s (standalone); 2.7369\,s (with TEF+IF); CPU overhead $\sim$4.90\% on RPi4; trust-level agreement management; stable operation &
Environmental, Economic, Social (trust) &
Selected via KVI contribution to trustworthiness and reduced downtime/OPEX (automation, zero-touch); low overhead and TRL 4--5 PoC validate KG-based selection for trust-aware E2E design \\
\hline
\textbf{Intent-based management (V.E)} &
Execution-time CDF: 2.69--4.16\,s, avg.\ $\sim$3.10\,s; stable non-RT management latency; supports SLA/service orchestration efficiency &
Environmental, Economic, Social &
Retained due to KVI alignment with OPEX reduction, improved downtimes, and operational stability/resilience; measured stability supports the KG-driven inclusion for sustainable operations \\
\hline
\textbf{Integration fabric (V.F)} &
Average latencies 1.66--3.78\,s; 100\% availability; consistent throughput; supports closed-loop coordination; enables service exposure and multi-domain operation &
Environmental, Economic, Social &
Selected due to KVI alignment (OPEX reduction, service uptime, safety via monitoring/data availability, exposure-driven value) and KPI evidence (availability/latency); PoC outcomes corroborate KG pruning \\
\hline
\end{tabular}%
} 
\end{table*}

\textcolor{black}{Table~\ref{tab:secV_concluding_validation} consolidates PoC findings and KVI mapping in Table \ref{tab:catmap} to validate KG-driven enabler selection. The results show that the enablers retained by KPI and KVI-driven pruning exhibit measurable KPI gains and KVI-relevant outcomes that align with environmental, economic, and social sustainability objectives.}

\section{CONCLUSION} \label{sec6}

\textcolor{black}{This article addresses an important gap: designing a 6G E2E system that considers KPI and KVI requirements of 6G use cases.} To do this, firstly a discussion on how the sustainability parameters, i.e., KVs and KVIs, are derived from the high-level human and planetary goals has been provided. Next, a novel framework based on Knowledge graphs for 6G E2E system design has been introduced. This novel framework formulates the 6G system design problem as a knowledge graph construction and pruning problem. It takes as its inputs the use case and its requirements, enablers and their features, design principles, and the parameters and constraints for the design of the 6G system. Subsequently, the graph is pruned based on the parameters and constraints to obtain a set of enablers that will form the 6G system for the use case under consideration. 

This process of graph construction and pruning was demonstrated on the cooperative mobile robots (COBOTS) use case, wherein the enabler selection methodology based on the KG framework selected 82 enablers from a set of more than 100 enablers. The methodology also demonstrated a novel approach for mapping enablers to key values (KVs). Specifically, for each KV there exist certain KVIs, which are subsequently mapped to technical requirements needed to be satisfied to achieve those KVI targets. Hence, in the proposed approach, the enablers under consideration are mapped to the technical requirements, which permits mapping of the enabler to the KVIs and KVs they satisfy. Lastly, proof-of-concept results for a subset of the selected enablers has been presented, which reinforces the selection done by the KG method. This is because, the results demonstrate a strong alignment of the selected enablers with the KPI and KVI requirements of the COBOTS use case, the system design principles as well as the maturity and criticality towards migration requirements. 

\textcolor{black}{As next steps, the proposed framework will be further extended to incorporate a more rigorous theoretical grounding of the KVI and enabler mapping as well as the associated weighting and pruning mechanisms. In addition, a more comprehensive quantitative evaluation against alternative design methodologies, such as KPI-only or rule-based approaches, will be carried out to further assess the benefits and tradeoffs of the proposed KPI+KVI-driven design paradigm.} Furthermore, given that a full KG with complete node and edge feature list is already built in this work, the concept of graph neural networks \cite{abadal2021} will be explored to generate node and edge embeddings. These can be then utilized to perform graph pruning as well as future dependency analysis through link prediction. The proposed approach will also be iteratively evolved in order to take into account any new upcoming technological enablers, use case requirements and parameters or constraints other than maturity and importance towards migration. 

It must be stated that the proposed KG framework can be easily utilized to design any complex system with 6G just as one example. Hence, as part of the future work a study on the performance of the KG framework for designing systems in areas of defense, industrial automation, backend IT and banking solutions, etc., will also be carried out. 

\section*{ACKNOWLEDGMENT}
The authors would like to acknowledge Ali Rezaki, M\aa rten Ericson, Nafiseh Mazloum and Lisa Lessing for their valuable insights and discussions during the Hexa-X-II project meetings.

\bibliographystyle{unsrt}
\bibliography{references}

\end{document}